\documentclass[twocolumn]{aastex61}

\usepackage{ulem}
\usepackage{soul}
\usepackage{multirow}
\usepackage{graphicx}
\usepackage{amsmath}
\usepackage{epsfig}
\usepackage{amssymb}
\usepackage{ulem}
\usepackage{hyperref}	

\newcommand{\spr}{$s$-process }
\newcommand{\sprn}{$s$~process }
\newcommand{\ipr}{$i$-process }
\newcommand{\iprn}{$i$~process }

\newcommand{\rpr}{$r$-process }
\newcommand{\rprn}{$r$~process }

\newcommand{\code}[1]{\texttt{#1}}

\newcommand{\mppnp}{\code{mppnp}}
\newcommand{\omegac}{\code{OMEGA}}
\newcommand{\sygma}{\code{SYGMA}}
\newcommand{\startrack}{\code{StarTrack}}
\newcommand{\mesa}{\code{MESA}}

\newcommand{\natlog}[2]{\ensuremath{#1\times 10^{#2}}} 
\newcommand{\unitstyle}[1]{\ensuremath{\mathrm{#1}}}
\newcommand{\Kelvin}{\unitstyle{K}}
\newcommand{\K}{\Kelvin}  
\newcommand{\minute}{\unitstyle{min}} 

\usepackage{xcolor}

\newcommand{\msun}{\ensuremath{\mathrm{M}_\odot}}
\newcommand{\rsun}{\ensuremath{\mathrm{R}_\odot}}

\submitjournal{ApJ}

\shorttitle{i-Process Contribution of Rapidly Accreting White Dwarfs}
\shortauthors{C\^ot\'e et al.}

\begin{document}

\title{i-process contribution of rapidly accreting white dwarfs to \\the solar composition of first-peak neutron-capture elements}

\correspondingauthor{Benoit C\^ot\'e}
\email{bcote@uvic.ca, benoit.cote@csfk.mta.hu}

\author[0000-0002-9986-8816]{Benoit C\^ot\'e}
\affil{Department of Physics and Astronomy, University of Victoria, Victoria, BC, V8W 2Y2, Canada}
\affiliation{Konkoly Observatory, Research Centre for Astronomy and Earth Sciences, Hungarian Academy of Sciences, Konkoly Thege Miklos ut 15-17, H-1121 Budapest, Hungary}
\affiliation{Joint Institute for Nuclear Astrophysics - Center for the Evolution of the Elements, USA}
\affiliation{NuGrid Collaboration, \href{http://nugridstars.org}{http://nugridstars.org}}

\author[0000-0001-6120-3264]{Pavel Denissenkov}
\affiliation{Department of Physics and Astronomy, University of Victoria, Victoria, BC, V8W 2Y2, Canada}
\affiliation{Joint Institute for Nuclear Astrophysics - Center for the Evolution of the Elements, USA}
\affiliation{NuGrid Collaboration, \href{http://nugridstars.org}{http://nugridstars.org}}

\author[0000-0001-8087-9278]{Falk Herwig}
\affiliation{Department of Physics and Astronomy, University of Victoria, Victoria, BC, V8W 2Y2, Canada}
\affiliation{Joint Institute for Nuclear Astrophysics - Center for the Evolution of the Elements, USA}
\affiliation{NuGrid Collaboration, \href{http://nugridstars.org}{http://nugridstars.org}}

\author[0000-0002-4794-6835]{Ashley J. Ruiter}
\affiliation{Research School of Astronomy and Astrophysics, Australian National University, Canberra, ACT 0200, Australia}
\affiliation{ARC Centre of Excellence for All-sky Astrophysics (CAASTRO)}
\affiliation{School of Physical, Environmental and Mathematical Sciences, University of New South Wales, Australian Defence Force Academy, Canberra, ACT 2600, Australia}

\author{Christian Ritter}
\affiliation{Keele University, Keele, Staffordshire ST5 5BG, United Kingdom}
\affiliation{Department of Physics and Astronomy, University of Victoria, Victoria, BC, V8W 2Y2, Canada}
\affiliation{Joint Institute for Nuclear Astrophysics - Center for the Evolution of the Elements, USA}
\affiliation{NuGrid Collaboration, \href{http://nugridstars.org}{http://nugridstars.org}}

\author{Marco Pignatari}
\affiliation{E.A. Milne Centre for Astrophysics, Department of Physics \& Mathematics, University of Hull, HU6 7RX, United Kingdom}
\affiliation{Joint Institute for Nuclear Astrophysics - Center for the Evolution of the Elements, USA}
\affiliation{NuGrid Collaboration, \href{http://nugridstars.org}{http://nugridstars.org}}

\author{Krzysztof Belczynski}
\affiliation{Nicolaus Copernicus Astronomical Center, Polish Academy of Sciences, ul. Bartycka 18, 00-716 Warsaw, Poland}



\begin{abstract}

Rapidly accreting white dwarfs (RAWDs) have been proposed as
contributors to the chemical evolution of heavy elements in the
Galaxy. Here, we test this scenario for the first time and
determine the contribution of RAWDs to the solar composition of
first-peak neutron-capture elements. We add the
metallicity-dependent contribution of RAWDs to the one-zone galactic
chemical evolution code \omegac\ according to RAWD rates from
binary stellar population models combined with metallicity-dependent
\ipr stellar yields calculated following the models of Denissenkov et al.~(2017).
With this approach we find  that the contribution
of RAWDs to the evolution of heavy elements in the Galaxy could be responsible for a
significant fraction of the solar composition of Kr, Rb, Sr, Y, Zr,
Nb, and Mo ranging from $2$ to $45\%$ depending on the element,
the enrichment history of the Galactic gas, and the total mass ejected
per RAWD. This contribution could explain the missing solar
Lighter Element Primary Process for some elements (e.g., Sr, Y, and
Zr). We do not overproduce any isotope relative to the solar
composition, but $^{96}$Zr is produced in a similar
amount. The \iprn produces efficiently the Mo stable isotopes
$^{95}$Mo and $^{97}$Mo. When nuclear reaction rate uncertainties
are combined with our GCE uncertainties, the upper limits for the
predicted RAWD contribution increase by a factor of $1.5-2$ for Rb,
Sr, Y, and Zr, and by 3.8 and 2.4 for Nb and Mo, respectively.
We discuss the implication of the RAWD stellar evolution properties on the single degenerate
Type~Ia supernova scenario.

\end{abstract}


\section{Introduction}
First-peak elements near Sr, Y, and Zr in the universe
have mainly been produced by the slow neutron-capture process ($s$~process) in 
asymptotic giant branch (AGB) stars (e.g., \citealt{1998ApJ...497..388G,2003ApJ...586.1305L,2004ApJ...601..864T,2005ARA&A..43..435H,2014ApJ...787...10B,2014PASA...31...30K,2015ApJ...801...53C,2016ApJ...827...30B}).
But these elements can also be synthesized by other stellar sources,
such as electron-capture supernovae (\citealt{2011ApJ...726L..15W}), the weak \sprn (e.g., \citealt{1990A&A...234..211P,1993ApJ...419..207R,2001ApJ...549.1085H,2007PrPNP..59..174H,2010ApJ...710.1557P,2016MNRAS.456.1803F})
and the strong \sprn (\citealt{2013ApJ...762...31P}) during the evolution of massive stars, neutrino-driven winds in core-collapse
supernovae (e.g., \citealt{2006PhRvL..96n2502F,2006ApJ...647.1323W,2013ApJ...770L..22W,2012ApJ...758....9N,2013JPhG...40a3201A,2014JPhG...41d4008M}), and neutrino-driven winds
following compact binary mergers (e.g., \citealt{2014MNRAS.443.3134P,2015MNRAS.448..541J,2015ApJ...813....2M}).

It is still unclear quantitatively to what
extent each of these sources has contributed to the chemical
evolution of the Galaxy in general and specifically to the
composition of the Sun.  \cite{2017ApJ...834L..10D} have shown that rapidly accreting
white dwarfs (RAWDs, see Section~\ref{sect_yields}) can also produce
first-peak elements via the intermediate neutron-capture process
($i$~process). Their calculations suggested that RAWDs
may be relevant for the chemical evolution of
elements between Ge and Mo.  The goal of the present paper is to
determine the contribution of RAWDs to the solar
composition in a galactic chemical evolution (GCE) model using
metallicity-dependent RAWD birthrates and \ipr yields.

GCE models calculate the
contribution of multiple stellar generations to the chemical evolution
of a galaxy (e.g., \citealt{1971ApJ...170..409T,1997ApJ...477..765C,2003PASA...20..401G,2013ARA&A..51..457N}). These models
ideally should take into account the formation time and initial
metallicity of all stellar populations.  Indeed, the various sources
of enrichment such as AGB stars, massive stars, Type~Ia supernovae
(SNe~Ia), compact binary mergers, and RAWDs, release their ejecta on
different timescales (e.g., \citealt{1979ApJ...229.1046T,2009ApJ...699.2026R,2012ApJ...759...52D}) and have
different chemical compositions depending on metallicity (e.g.,
\citealt{1998A&A...334..505P,2004ApJ...608..405C,2006ApJ...653.1145K,2015ApJS..219...40C,2016ApJ...825...26K,2016ApJS..225...24P}).  In addition, the metallicity
can affect the rate at which an enrichment source is releasing its
ejecta (see Section~\ref{sect_pop_synth}).  Therefore, when
considering enrichment sources with metallicity-dependent
  properties, as it is the case for RAWDs, it is necessary to follow such contributions  in
a GCE model.
 
The paper is organized as follows. In Section~\ref{sect_yields}, we present our 
\ipr nucleosynthetic yields calculation and discuss their metallicity dependence.
In Section~\ref{sect_pop_synth}, we describe the population synthesis model used
to derive the time- and metallicity-dependent rates for RAWDs. Our galactic
chemical evolution model for the Milky Way is described in Section~\ref{sect_MW_model}
and results are shown in Section~\ref{sect_results}. A discussion is provided
in Section~\ref{sect_disc} on various sources of uncertainty and on the implication 
of our results for the solar Lighter Element Primary Process (LEPP). In Section~\ref{sect_conc}, we present
our conclusions.

\section{RAWD i-process Yields} \label{sect_yields}
RAWDs are carbon-oxygen (CO) or oxygen-neon white-dwarf primary stars in a close binary system, with a main-sequence, subgiant, red-giant branch or asymptotic giant branch
secondary component. The RAWD accretes H-rich material from the  companion rapidly, at mass accretion rates around
$\dot{M}_\mathrm{acc}\sim 10^{-7}\,$M$_\odot\ \mathrm{yr}^{-1}$ \citep[e.g.][]{2007ApJ...663.1269N} and 
the accreted H burns steadily in a shell leaving behind an accumulating layer of He ash. 
At lower rates, the accreted H shell will periodically experience mild thermal flashes that will become stronger as the accretion rate
decreases, eventually leading to nova events. At higher rates, the accreted H shell will expand forming a red-giant envelope
\citep[e.g.,][and references therein]{2013ApJ...777..136W,2013ApJ...778L..32M}.

The accumulating He shell eventually experiences a He-shell flash \citep{cassisi:98}, a cycle which is then repeated for a few dozen or so times \citep{2017ApJ...834L..10D}.
The fact that stable H-shell burning is periodically interrupted by He-shell flashes is, of course, familiar from thermal pulses that occur for all core masses eventually in AGB stellar models.
A  post-AGB star can also experience a  very late thermal pulse (VLTP) on the WD cooling track \citep{2001Ap&SS.275...15H}.

A high energy output during the He-shell flash triggers convection, and in the VLTP case the upper convection boundary can approach the surrounding stable H-rich envelope and eventually mix that H with the products of He burning. The protons are advected downward in the convective He-burning shell where the
$^{12}$C abundance is $\approx 20 - 40\%$. The ingested protons are rapidly consumed when reaching $T\approx \natlog{1.5}{8} \K$ via the reaction $^{12}$C(p,$\gamma)^{13}$N.  Unstable $^{13}$N
with the half-life of $9.97 \minute$ decays into $^{13}$C while being transported by convection toward the bottom of the He shell, where
neutrons are released in the reaction $^{13}$C($\alpha$,n)$^{16}$O. Depending on its parameters, the neutron density
in this process can reach a value of $N_\mathrm{n}\sim 10^{15}\ \mathrm{cm}^{-3}$ \citep{1986MNRAS.223..683M}, which is intermediate 
between the values typical for the $s$ and $r$ processes, and thus termed \iprn \citep{1977ApJ...212..149C}.

The surface abundances of heavy elements, including the first-peak \spr elements Rb, Sr, Y and Zr, measured by \cite{1999A&A...343..507A}
in the post-AGB star Sakurai's object (V4334 Sagittarii) and their interpretation by \cite{2011ApJ...727...89H} provided the first strong evidence of
the \ipr nucleosynthesis in VLTP stars. Because a single post-AGB star undergoes just one He-shell flash, during which only 
a small amount of $i$-processed mass ($\Delta M_\mathrm{He} < 0.03\,$M$_\odot$) is ejected, the VLTPs should not contribute much to the GCE of heavy elements.
However, this situation may change if the post-AGB star is a RAWD. The key question in this case is -- will the RAWD eject a significant fraction of
the $i$-processed He-shell material after each of its TPs?

To answer this, \citet{2017ApJ...834L..10D} have simulated multiple He-shell flashes on RAWDs with solar initial chemical composition 
 [Fe/H]\footnote{We use the standard spectroscopic notation [A/B] = $\log_{10}(N(\mathrm{A})/N(\mathrm{B})) -
\log_{10}(N_\odot(\mathrm{A})/N_\odot(\mathrm{B}))$, where $N(\mathrm{A})$ and
$N(\mathrm{B})$ are the mass fractions or number densities of the nuclides A
and B.}=\,0. Accordingly thermally-pulsing RAWDs lose $\geq 90\%$ of their accumulated and then $i$-process-element enriched He shells. The resulting He-retention efficiencies, representing a ratio of the He-shell mass left on the RAWD 
to the ejected mass, are consequently $\eta_\mathrm{He}\la 10$\%. 
After each He-shell flash the envelope of a RAWD expands and remains so until almost
the entire mass accumulated between two consecutive TPs is ejected either by
the super-Eddington luminosity wind mass loss or by Roche-lobe overflow. 

\begin{figure}
\center
\includegraphics[width=3.3in]{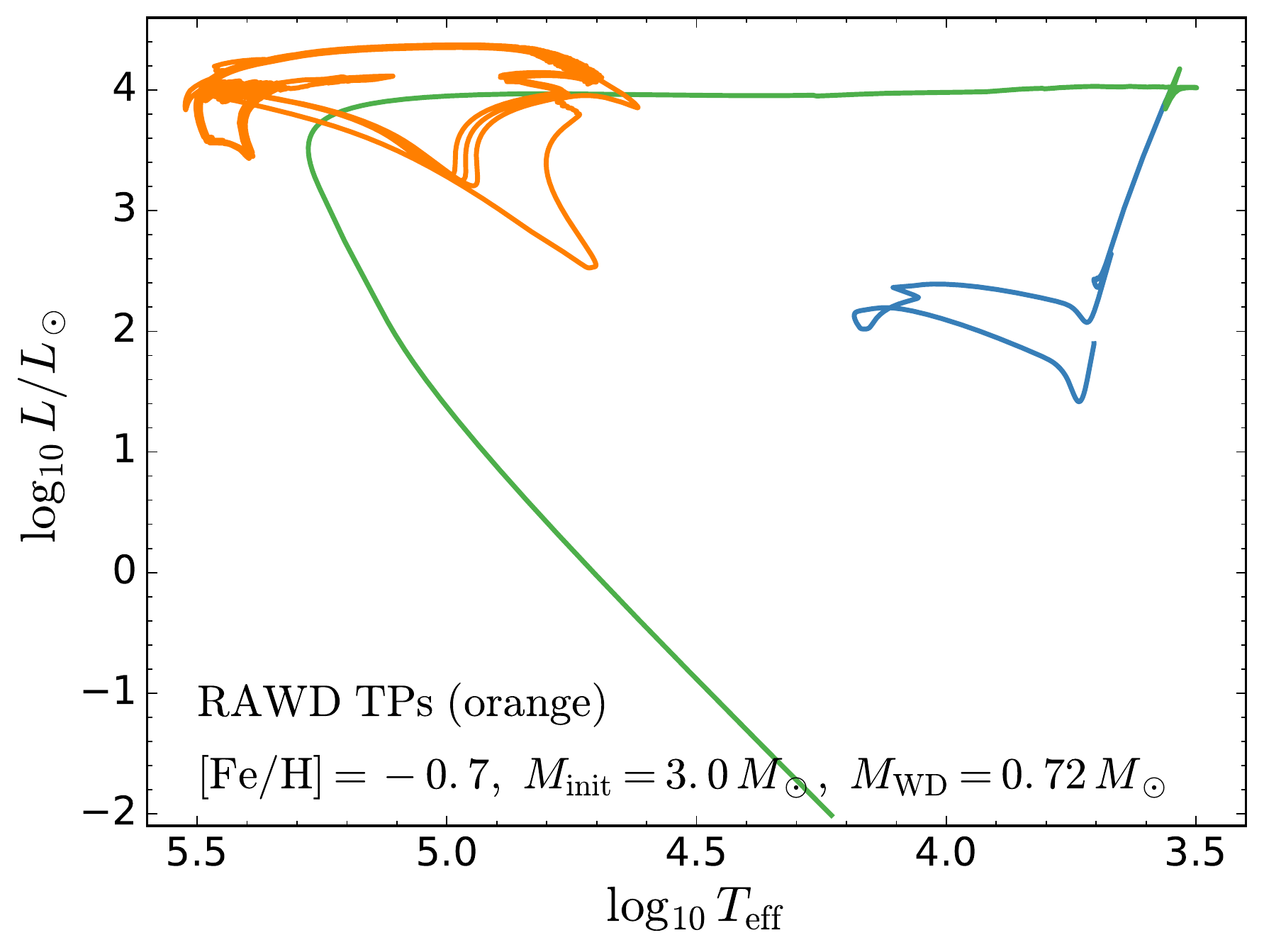}
\caption{An example of stellar evolution tracks from our new RAWD simulations. The blue track shows the evolution of
an initially $3\,$M$_\odot$ model from the pre-MS phase through to its first He-shell pulse on the AGB. The model
is then forced to lose almost its entire H-rich envelope in a presumably common-envelope event and the remaining $0.72\,$M$_\odot$
CO core moves towards the WD cooling track (the green curve). The orange track with multiple loops
shows the RAWD evolution that consists of H-accreting phase, followed by a He-shell thermal pulse (TP), envelope expansion, its loss
via the Roche-lobe overflow, and return of the model to the accretion phase.}
\label{fig_RAWD_TP}
\end{figure}

We have extended the RAWD simulations to the following lower initial chemical compositions:
[Fe/H]\,$=0.0,\ -0.7,\ -1.1,\ -1.55$, and $-2.3$. We adopt the \cite{asplund09} solar abundance distribution which implies
the heavy-element mass fractions $Z_\mathrm{met}=0.014,\ 0.0054,\ 0.0021,\ 0.00076$, and $0.00014$, respectively\footnote{Throughout
this paper, we use $Z_\mathrm{met}$ for metallicity in mass fraction in order to avoid confusion with $Z$, the elemental charge number.}. Details of our new RAWD simulations
will be presented elsewhere. 
Here, we are using only the \ipr yields calculated for CO WD masses that are all close to $0.7\,$M$_\odot$. 
Figure~\ref{fig_RAWD_TP} shows as an example the  stellar evolution track for [Fe/H]\,$=-0.7$ computed with the \mesa\ code \citep[revision 7624][]{2013ApJS..208....4P}. The blue curve is a track of an initially $3\,$M$_\odot$ star from the pre-MS
evolutionary phase through to its first He-shell flash on the AGB. After that, the model star is forced to lose its envelope, as if 
a common-envelope event occurred to it and, as a result, it leaves the AGB and moves to the WD cooling track (the green curve). 
The accretion of H-rich material begins after the $0.72\,$M$_\odot$ CO WD has cooled down to $\log_{10}L/L_\odot = -2$. We start with a slow accretion,
$\dot{M}_\mathrm{acc}\sim 10^{-8}-10^{-9}\,$M$_\odot\,\mathrm{yr}^{-1}$, more typical for novae, to ensure numerical convergence, and
we switch to the rapid accretion at a later time.
The orange curve shows the multiple loops that the evolutionary track of the RAWD makes
when it experiences He-shell flashes followed by its expansion, mass loss due to the Roche-lobe overflow, and return to the mass-accreting phase. 
The pathway to RAWDs adopted for our yield calculations (Figure~\ref{fig_RAWD_TP}) is also present in our binary population synthesis models.

\begin{figure}
\center
\includegraphics[width=3.3in]{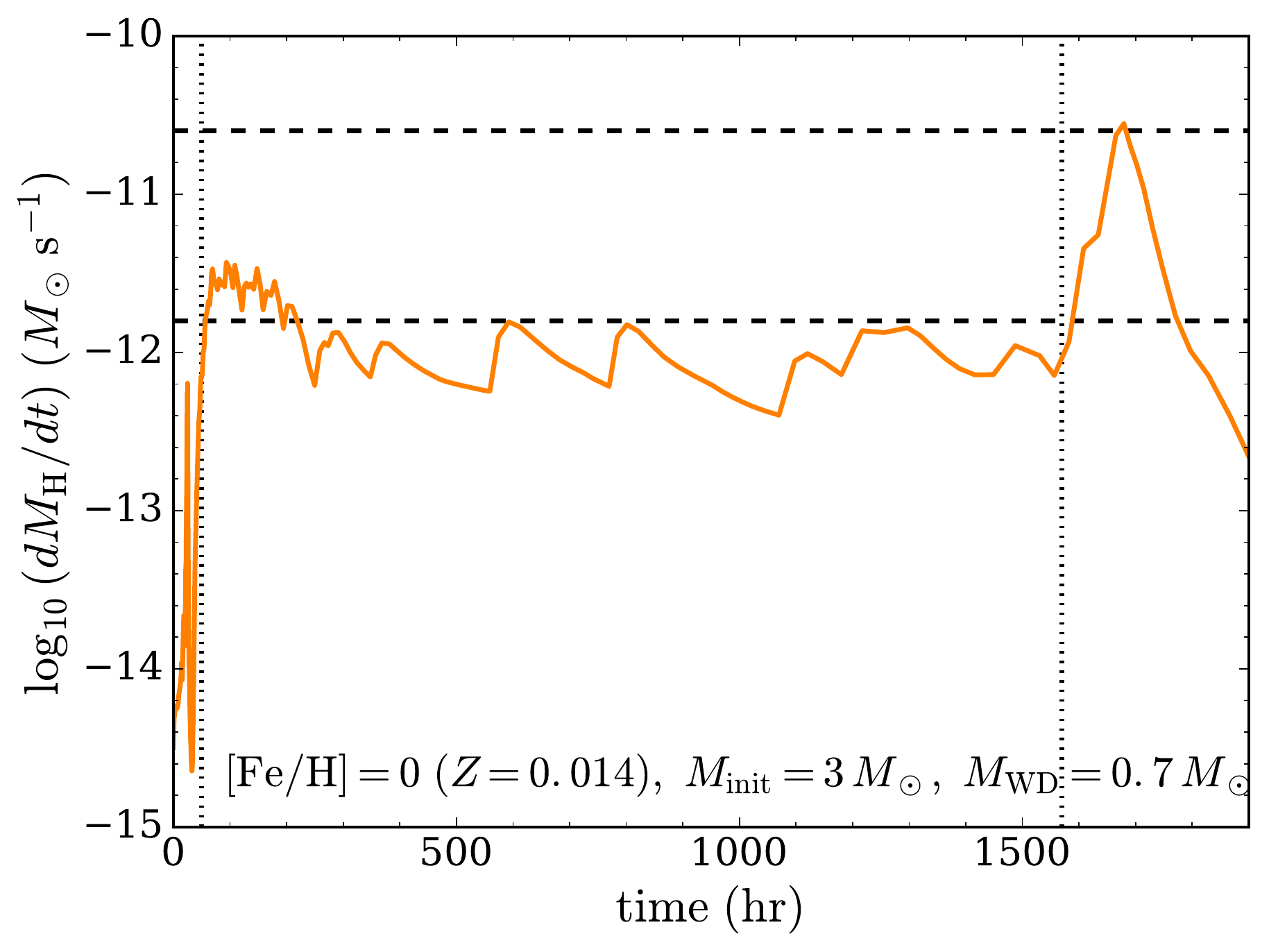}
\caption{Rate and duration of H ingestion in the solar-metallicity RAWD model estimated from its H-burning luminosity.
Note the final short phase with an increased rate of H ingestion that is typical for near-solar metallicity RAWD models.}
\label{fig_MHdot}
\end{figure}

For the \iprn to be activated, the He-shell convection has to ingest some H from its surrounding H-rich envelope.
In our 1D stellar evolution models of RAWDs this happens even if no convective overshooting is assumed
\citep[see Figure~2 in][]{2017ApJ...834L..10D}.
When convective boundary mixing at the top boundary of the pulse-driven convection zone is included according to the exponentially decaying diffusive model with an efficiency $f=0.01$ as recommended by
\cite{2007ASPC..378...43H}, the 1D RAWD models have H-ingestion rates 
of $\dot{M}_\mathrm{H}\sim 10^{-11}-10^{-12}\,$M$_\odot\,\mathrm{s}^{-1}$, as estimated from their H-burning luminosities.
These are consistent within a factor of 2 to those obtained in 3D hydro simulations of H ingestion by
He-flash convection, using the convective He-shell structure and He-burning luminosities from our RAWD models (R. Andrassy, priv.\ com.). 
These values of $\dot{M}_\mathrm{H}$ have been used in our post-processing 
nucleosynthesis computations of the \iprn in RAWDs.
We have carried out these computations using the multizone frame 
\mppnp\ of the NuGrid code \citep{2016ApJS..225...24P}.
Durations of the H ingestion events have been estimated from our 1D RAWD models. In single post-AGB stars
VLTPs induce a violent H ingestion that has a higher mass ingestion rate ($\dot{M}_\mathrm{H}\sim 10^{-10}\,$M$_\odot\,\mathrm{s}^{-1}$) 
than in the preceding thermal pulse evolution. This high ingestion rate is only maintained for a short time  \citep[hundreds of minutes, ][]{2011ApJ...727...89H}, 
while in RAWDs H ingestion is usually 10 to 100 times slower, not accompanied by violent H burning or major perturbations of the convective structure of the He-shell,
and it lasts tens of days.  In the case of [Fe/H]\,$=0$, such a long-lasting gentle H ingestion is followed by a much shorter and stronger H-ingestion event
that resembles the violent H ingestion after a VLTP and that terminates the whole H-ingestion process (Figure~\ref{fig_MHdot}). 
We take this into account in our post-processing nucleosynthesis computations by changing $\dot{M}_\mathrm{H}$ appropriately 
in our solar-metallicity RAWD models.

\begin{figure}
\center
\includegraphics[width=3.3in]{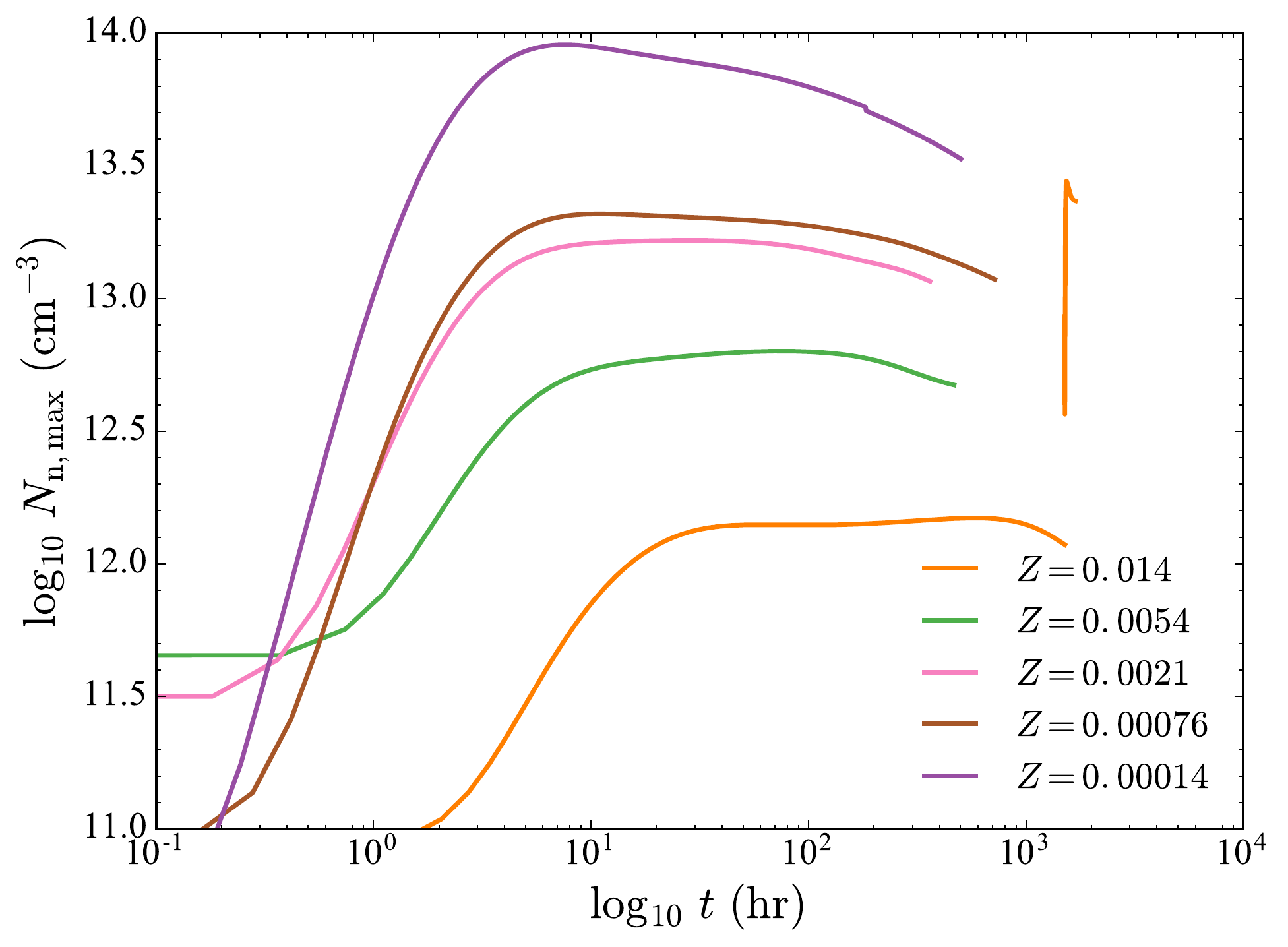}
\caption{Maximum neutron densities in the He convective zones from the post-processing computations of
the \ipr nucleosynthesis in our RAWD models. The almost vertical part of the orange curve corresponds
to the final fast H-ingestion event in the solar-metallicity RAWD model that is seen in Figure~\ref{fig_MHdot}.}
\label{fig_iRAWD_N}
\end{figure}

\begin{figure}
\center
\includegraphics[width=3.3in]{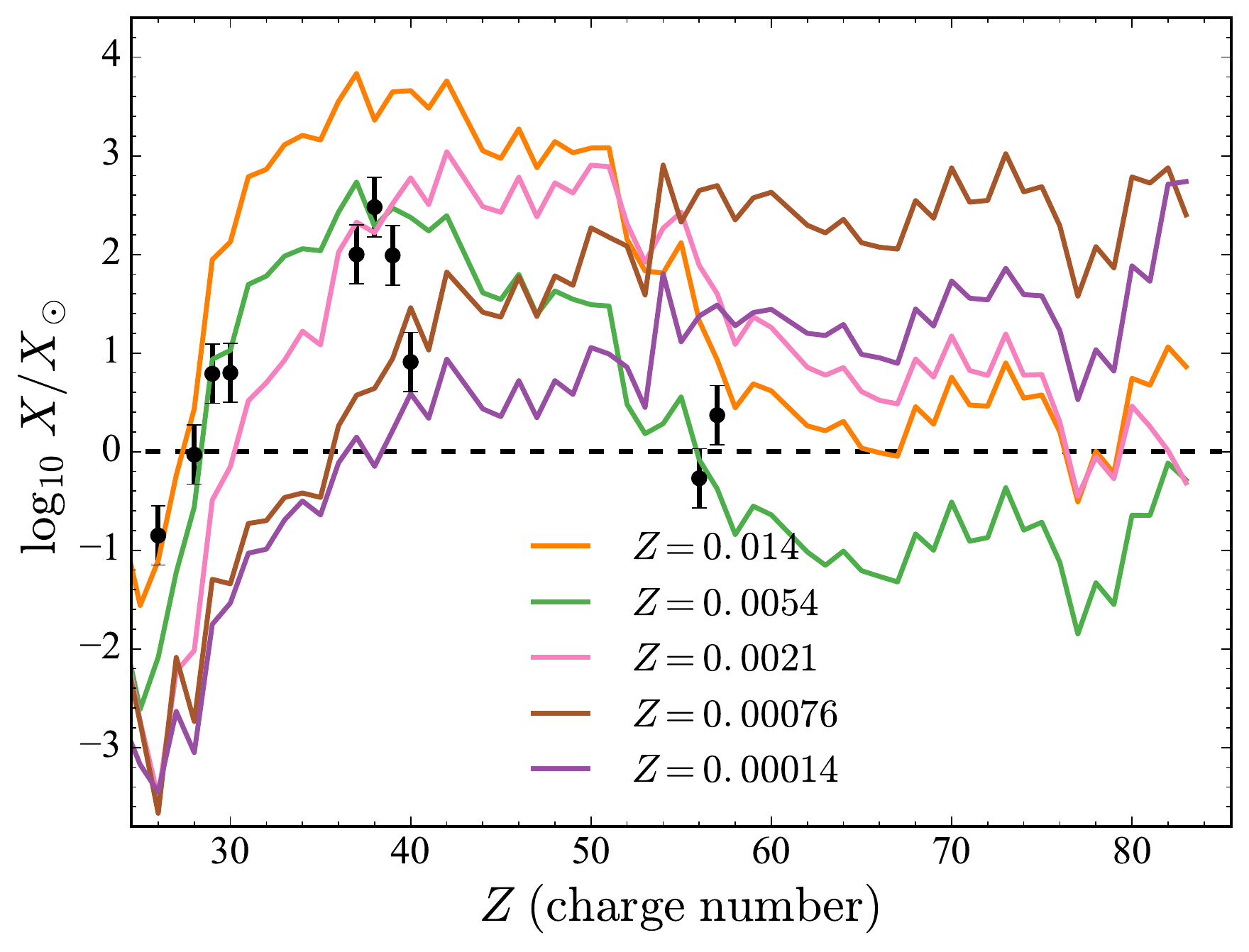}
\caption{Distributions of element yields from the post-processing computations of the \ipr
nucleosynthesis in our RAWD models. The black circles with error bars are surface abundances
in Sakurai's object measured by \cite{1999A&A...343..507A}. Note the high abundances of the first-peak elements
in the nearly-solar metallicity models.}
\label{fig_iRAWD_Y}
\end{figure}

Figure~\ref{fig_iRAWD_N} shows maximum neutron densities in the convective He shells of our post-processed RAWD models
as a function of time. The orange curve consists of two parts, the second, almost vertical one, corresponding to the final
strong H ingestion event that we have revealed in the solar-metallicity model (Figure~\ref{fig_MHdot}). 
The peak value of $N_\mathrm{n,\,max}$ increases when the metallicity decreases
because of a decreasing total mass fraction of the neutron-capture seeds. 
This results in a shift of the final distribution of \ipr
yields towards heavy elements (Figure~\ref{fig_iRAWD_Y}). 
However, for the main topic of this work it is more important to
comment on the RAWD yields of the first-peak elements with the charge number around $40$. The black circles with error bars
in Figure~\ref{fig_iRAWD_Y} show the surface abundances in Sakurai's object measured by \cite{1999A&A...343..507A}. In 
terms of abundance distribution, the RAWD yields at near-solar metallicity contain similar or even higher amounts of first-peak elements
compared to Sakurai's object. Given that RAWDs can potentially undergo tens of He-shell flashes with low He-shell
mass retention efficiencies, they can indeed be important contributors to the GCE evolution of these elements, as
was originally proposed by \cite{2017ApJ...834L..10D}.

Isotopes with large neutron-capture cross sections that act as neutron poisons are all
automatically included in our nucleosynthesis computations. We begin with the abundance distributions 
in the He convective zones obtained from the solar-scaled abundances processed through complete H burning followed by partial
He burning.  The ingested material has the same initial solar-scaled chemical composition, and the NuGrid codes that we use
take into account all the relevant reactions ($\sim$\,$14{,}000$ reactions for the models presented in Figure~\ref{fig_iRAWD_Y} and
$\sim$\,$61{,}000$ for test runs).

The RAWD \ipr elemental yields from Figure~\ref{fig_iRAWD_Y} supplemented by 
their corresponding isotopic yields are used as input data for the GCE model described in Section~\ref{sect_MW_model}. These yields
represent decayed elemental and isotopic abundances mass-averaged over convective He shells.

\section{Population Synthesis Model} \label{sect_pop_synth}
Our binary star populations that give rise to the RAWD systems are simulated with the \startrack\ rapid binary evolution population synthesis code \citep{2002ApJ...572..407B,2008ApJS..174..223B}. We simulate stellar populations from the zero-age main sequence (ZAMS) up to a Hubble time. Assuming a binary fraction of 70\%, all stars are born in a starburst at $t=0$, and later convolved with the appropriately-chosen star formation history and star formation efficiency (see Section~\ref{sect_MW_model}). Our four populations are evolved using four different ZAMS metallicities: $Z_\mathrm{met} = 0.02$, 0.002, 0.001, and 0.0001. 
The effect of initial metallicity on the binary evolution, and thus on the RAWD birthrates, is discussed in Section~\ref{sect_effect_Z_RAWDs}. 

\begin{figure}
\center
\includegraphics[width=3.3in]{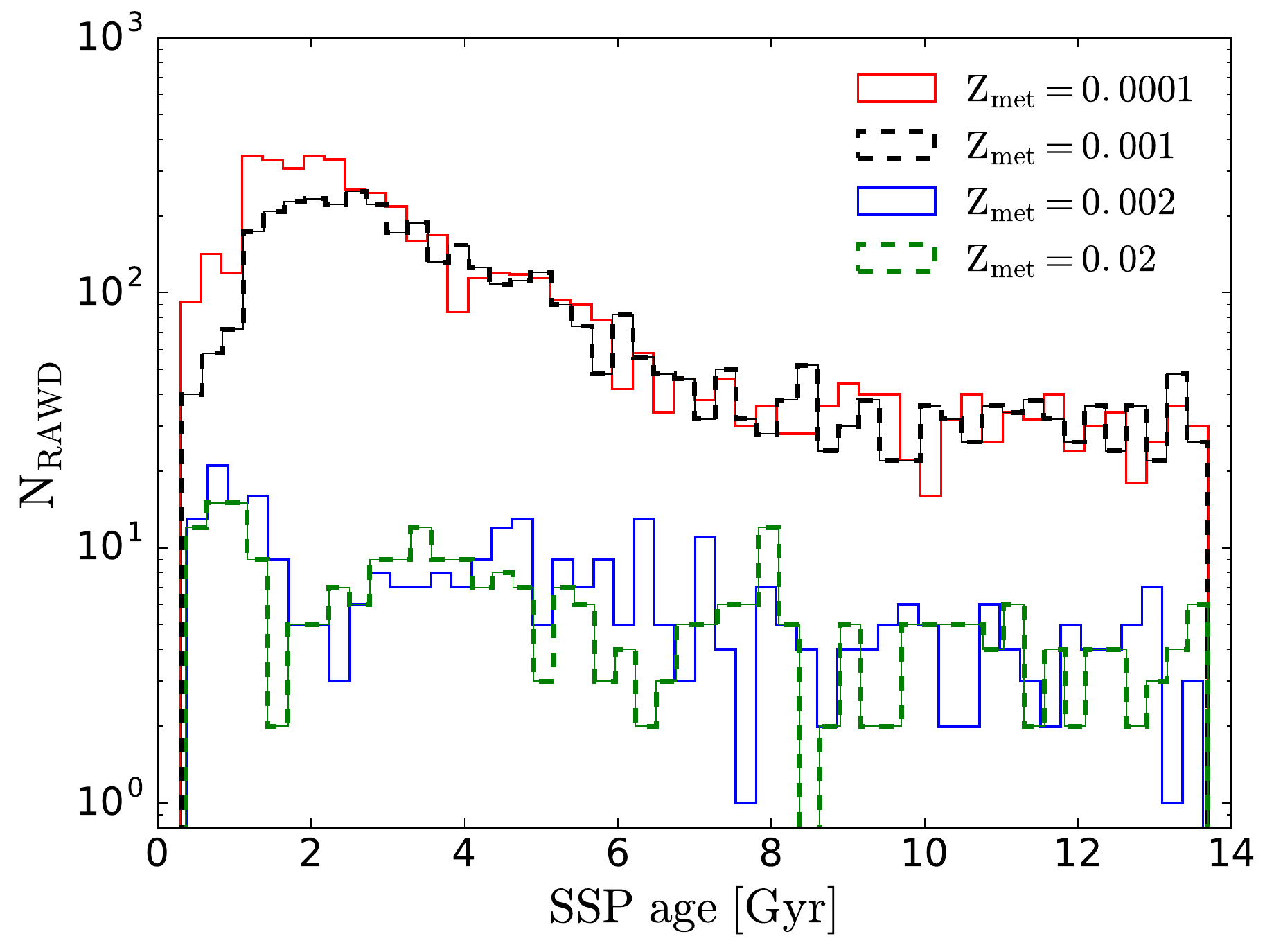}
\caption{Number of rapidly accreting white dwarfs (RAWDs) occurring in a simple stellar population (SSP) as a function of time, split into 50 bins. Different colors represent population synthesis predictions at different metallicity $Z_\mathrm{met}$ (see Section~\ref{sect_pop_synth}). Each SSP has a total stellar mass of $3.2\times10^6$\,$M_\sun$, which is formed instantaneously. In total, $1.58\times10^{-3}$, $1.26\times10^{-3}$, $1.0\times10^{-4}$, and $8.9\times10^{-5}$ RAWD event occurs per unit of stellar mass formed at $Z_\mathrm{met}=0.0001$, 0.001, 0.002, and 0.02, respectively.  We refer to Annexe~\ref{annexe_A} for a discussion of the sharp transition between $Z_\mathrm{met}=0.001$ and 0.002.}
\label{fig_rawd_rates}
\end{figure}

Initial ZAMS star masses are drawn from the 3-component power-law initial mass function of \citet{1993MNRAS.262..545K} with $\alpha_{1} = -1.3$, $\alpha_{2} = -2.2$, $\alpha_{3} = -2.35$.  The initially more massive star ($M_1$) and its companion ($M_2$) are chosen within the mass range of $0.8 - 100.0$ and $0.5 - 100$\,\msun, respectively\footnote{\startrack\ follows all type of binary systems including low-mass and massive stars, but only the ones involving white dwarfs are relevant for this study.}.  
$M_1$ is drawn directly from the probability distribution function given by our chosen IMF while $M_2$ is calculated by randomly picking a mass ratio $M_2/M_1$ between 0 and 1. \cite[][but see also \citealt{2017ApJS..230...15M}]{1991A&A...248..485D,2014A&A...562A..14T}. For simplicity, we assume circular orbits from the ZAMS and flat orbital separations (in the logarithm) from $2 \times $($R_1 + R_2$) to $10^{5}$\,\rsun\ (standard prescription).

Interacting binary stars undergo at least one common envelope (CE) phase over the course of their evolution. Though this phase is extremely important in bringing two stars close enough to one another to undergo mass transfer, it is one of the most poorly-understood processes in stellar astrophysics (see Section~\ref{sect_disc_pop}). 
In population synthesis studies, the CE phase cannot be explicitly calculated but must be parametrized in some way. A common approach is to equate the binding energy of the envelope of the mass-losing star, $E_{\rm bind} = G M_{\rm core} M_{\rm env} R^{-1} \lambda^{-1}$ (see below), with the orbital energy of the binary system.
The envelope will then be expelled from the system at the expense of the binary's orbital energy, which causes the orbital size to decrease, often drastically. 
We adopt the ``standard" common envelope formalism employing energy balance \citep{1984ApJ...277..355W} that is often used in binary population synthesis codes with $\alpha_{\rm CE} \times \lambda = 1$ \citep[see][]{2009ApJ...699.2026R}. Here, $\alpha_{\rm CE}$ is the fraction of orbital energy that is used to eject the envelope of the mass-losing star, and $\lambda$ is the binding energy parameter.

We consider a sub-population of our accreting white dwarfs to contribute to the RAWD population. Specifically, any CO WD with a mass $\ge 0.6$ \msun\ that accretes from any hydrogen-rich star at a rate $\ge3.066\times10^{-7} [(M_{\rm WD accretor}/\msun) - 0.5357$] \msun\ yr$^{-1}$ \citep[][see their Figure 4]{2007ApJ...663.1269N} is considered to be a RAWD in our models. 
For this study, unlike in previous studies \citep[e.g.][]{2009ApJ...699.2026R}, we artificially suppress hydrogen accumulation on the WD, as found in \citet{2017ApJ...834L..10D}. The implications of this for other sources, such as Type Ia supernovae (SNe~Ia), are discussed in Section~\ref{sect_sds}. 
The time (from star formation) when these accretion criteria are satisfied is considered to be the RAWD birth time (its ``delay time"). The delay time distribution (DTD) functions for the 4 RAWD populations are shown in Figure~\ref{fig_rawd_rates} and set the enrichment timescale of \ipr element that are implemented in our GCE model.

\section{Milky Way Model} \label{sect_MW_model}
In this section we briefly describe our galactic chemical evolution model
and compare its output properties with the Milky Way.

\subsection{Galactic Chemical Evolution Code}
We use the one-zone chemical evolution code \omegac\ described in \cite{2017ApJ...835..128C}, which is
available on GitHub as part of the open-source NuGrid Python Chemical Evolution Environment
(NuPyCEE,\footnote{\url{http://nugrid.github.io/NuPyCEE}} version 2.0).
From an input star formation history, which is decreasing with time in our case, the code
follows the contribution of several simple stellar populations (SSPs) to the overall stellar
ejecta by keeping track of the age, initial metallicity, and initial mass of each SSP.
\omegac\ uses the uniform-mixing approximation and accounts for galactic
outflows and primordial inflows. The rate of inflow at each timestep 
is automatically adjusted to sustain the input star formation rate. Our code offers
a variety of parametrization options for outflows and star formation efficiencies. But in this work,
we use the option described in \cite{2016MNRAS.463.3755C} which allows to control the early chemical evolution
of the galactic gas independently of its final properties. As seen in Section~\ref{sect_MW_prop},
this enables us to explore different chemical evolution paths to reach solar composition
and to quantify the confidence levels of the predicted contribution of RAWDs.

We use the NuGrid Set1 extension stellar yields (\citealt{2017arXiv170908677R})
for asymptotic giant branch (AGB) stars and massive stars including core-collapse supernova nucleosynthesis (see \citealt{2016ApJS..225...24P}).
Stellar models are provided at five metallicities from 0.0001 to 0.02 in mass fraction.
We also use the yields of \cite{2010ApJ...724..341H} for zero-metallicity stars and the W7 model of \cite{1999ApJS..125..439I} for 
SN~Ia yields. We use the stellar initial mass function of \cite{2001MNRAS.322..231K} for all
stellar populations at all metallicities. However, the choice of stellar yields is not particularly 
important for this work since we are only interested in the evolution of $Z_\mathrm{met}$, 
the overall gas metallicity (see Section~\ref{sect_MW_prop}).

\begin{deluxetable}{lcc}
\tablewidth{0pc}
\tablecaption{Properties of our galaxy model (\omegac) at the end of the simulation compared to current disk properties of the Milky Way taken from Table~1 in \citet[K15]{Kubryk15}. SFR, CC~SN, and SN~Ia stand for star formation rate, core-collapse supernova, and Type~Ia supernova. \label{tab_MW_prop}}
\tablehead{ \colhead{Quantity} & \colhead{\omegac} & \colhead{Milky Way (K15)}}
\startdata
Stellar mass [10$^{10}$\,$M_\odot$] & 5.0 & 3 -- 4 \\
Gas mass [10$^{9}$\,$M_\odot$] & 9.1 & $8.1\pm4.5$ \\
SFR [$M_\odot$ yr$^{-1}$] & 2.5 & 0.65 -- 3 \\
Inflow rate [$M_\odot$ yr$^{-1}$] & 1.4 & 0.6 -- 1.6 \\
CC~SN rate [per 100\,yr] & 2.5 & $2\pm1$ \\
SN~Ia rate [per 100\,yr] & 0.3 & $0.4\pm0.2$ \\
\enddata
\end{deluxetable}

We refer to \cite{2017ApJ...835..128C} for more information on \omegac\
and to \cite{ritter_sygma} for more information on the implementation of SSPs and SNe~Ia.

\subsection{RAWD Implementation} \label{sect_rawd_impl}
The contribution of RAWDs has been implemented in our SSP module
\sygma\ \citep{ritter_sygma}, which is called at each timestep by \omegac.
Because the gas metallicity increases continuously in our
one-zone galaxy model, each formed SSP has a unique metallicity
and thus has a unique set of \ipr yields and DTD function for their RAWDs
population. The yields are interpolated in the log-log space in order to represent the
initial metallicity of the stars.  The DTD functions are also interpolated to provide
a continuous evolution of RAWD rates as a function of galactic age. The 
total number of RAWD events in an SSP
depends on its total mass and on the normalization of its interpolated DTD function.
At a given timestep in our simulation, the overall RAWD ejecta is calculated by
summing the contribution of all existing SSPs and by keeping track of their 
specific age, mass, and unique set of interpolated \ipr yields and DTD function.

Each RAWD event is assumed to eject between 0.5 and 1\,M$_\odot$ of material. 
Our binary population synthesis simulations find $0.86\,M_\odot$ and $1.2\,M_\odot$ for the mean masses of the RAWD and its donor.
The isotropic re-emission approximation \citep[see Section 3.3.3 in][and references therein]{postnov:14}, 
that is appropriate for our RAWD binary models, provides a stable mass transfer for the accretor to donor mass ratio $q\la q_\mathrm{crit}\approx 1$.
This means that our RAWD models with the masses $\sim 0.7\,M_\odot$ should be able to stably accrete 
up to $\sim 0.5\,M_\odot$ from their $1.2\,M_\odot$ companion.
The $0.86\,M_\odot$ RAWDs would accrete $\sim 0.34\,M_\odot$. 
We think that our estimates of the total ejected mass have a factor of $\sim 2$ uncertainty.
The accretion itself usually takes a few Myr for $q$ to reach its critical value.

\begin{figure}
\center
\includegraphics[width=3.3in]{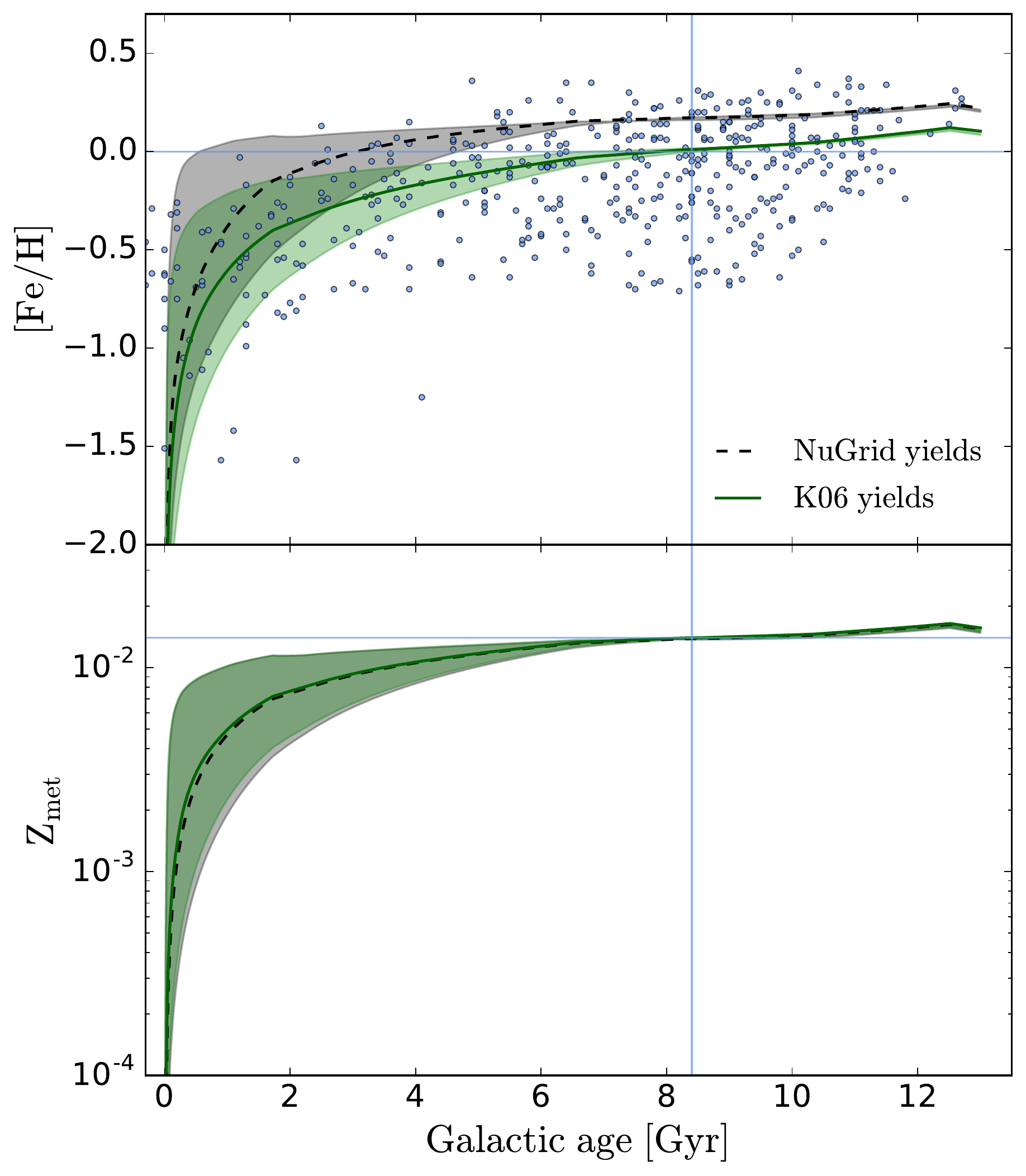}
\caption{Evolution of [Fe/H] (upper panel) and metallicity $Z_\mathrm{met}$ in mass fraction (lower panel) as a function of Galactic age. The black dashed and green solid lines represent our numerical predictions using NuGrid (\protect\citealt{2017arXiv170908677R}) and \protect\citet[K06]{2006ApJ...653.1145K} yields. For each set of stellar yields, the shaded area shows the different chemical evolution paths that can be predicted by our one-zone model at early times. These variations are used to provide confidence levels in the predicted contribution of rapidly accreting white dwarfs (see Section~\ref{sect_results}). The horizontal blue lines show the solar metallicity (\citealt{asplund09}).  The vertical blue line going across the two panels show the time at which the Sun is assumed to form in our model, 4.6\,Gyr \citep{2017GeCoA.201..345C} before the end of the 13\,Gyr-long simulation. Observational data was taken from \protect\citet[blue dots]{2014A&A...562A..71B}. Error bars for [Fe/H] data are about 0.05\,dex while the ones for Galactic age data can reach several Gyr. We reversed the time axis in the data so that the shortest \textit{look-back} time found in \protect\cite{2014A&A...562A..71B} corresponds to the end of our simulation. We did not include data with large uncertainties (grey dots in their Figure~21).}
\label{fig_age_met}
\end{figure}

\subsection{Milky Way Properties} \label{sect_MW_prop}
The focus of this paper is the chemical composition of the Galactic gas when the Sun forms.
We have tuned our chemical evolution model to ensure that the gas reaches solar
metallicity ($Z_{\mathrm{met},\odot}=0.014$, \citealt{2009LanB...4B...44L}) 4.6\,Gyr before
the end of the simulation, which lasts for 13\,Gyr.  We also tuned our model
to roughly reproduce the current observed properties of the Milky Way (see Table~\ref{tab_MW_prop}).
The upper panel of Figure~\ref{fig_age_met} shows the predicted evolution of [Fe/H] as a function of Galactic age.  
The dashed black and green solid lines represent our fiducial predictions using different sets of stellar yields. The shaded areas highlight the different chemical evolution paths to reach the Sun with our model.  These different paths are use to test the sensitivity or our results (see Section~\ref{sect_results}).

As described in \cite{2016MNRAS.463.3755C}, we can modify the gas content at early times (which modifies the metal concentration) without modifying the final properties of our galaxy model and the overall metallicity from which the Sun forms (lower-panel of Figure~\ref{fig_age_met}).  Because the contribution of SNe~Ia in the Milky Way should appear near [Fe/H]~$\sim-1$ (e.g., \citealt{1986A&A...154..279M,2001ApJ...554.1044C}), the lower limit for the evolution of [Fe/H] was chosen so that a value of $-1$ is reached at most after 1\,Gyr of evolution. This represents a comfortable lower limit given the prompt nature of SNe~Ia (e.g., \citealt{2005A&A...433..807M,2011MNRAS.412.1473L}) and their minimum delay times of ${\sim}10^8$ Myr (e.g, \citealt{2011MNRAS.417..408R,2017ApJ...834...15H}).

The choice of stellar yields for massive stars affects the scaling of [Fe/H]. Our SSPs tend to eject more Fe with NuGrid yields compared to when we use the ones found in \cite{2006ApJ...653.1145K} (see also \citealt{2017arXiv171205686P}). The choice of stellar yields, however, does not significantly impact the overall metallicity evolution in the Galactic gas (lower-panel of Figure~\ref{fig_age_met}). Because the goal of this paper is to quantify the contribution of RAWDs to the solar composition, our results are insensitive to the adopted stellar yields, since the predicted RAWD ejecta only depends on the overall metallicity, and not on its elemental composition.
\\
\\
\\
\section{Results} \label{sect_results}
In the following sections, we describe our predicted Galactic
RAWD rates and their contribution to the elemental and isotopic 
compositions of the Sun.

\subsection{RAWD Rates} \label{sect_rawd_rates}
\begin{figure}
\center
\includegraphics[width=3.3in]{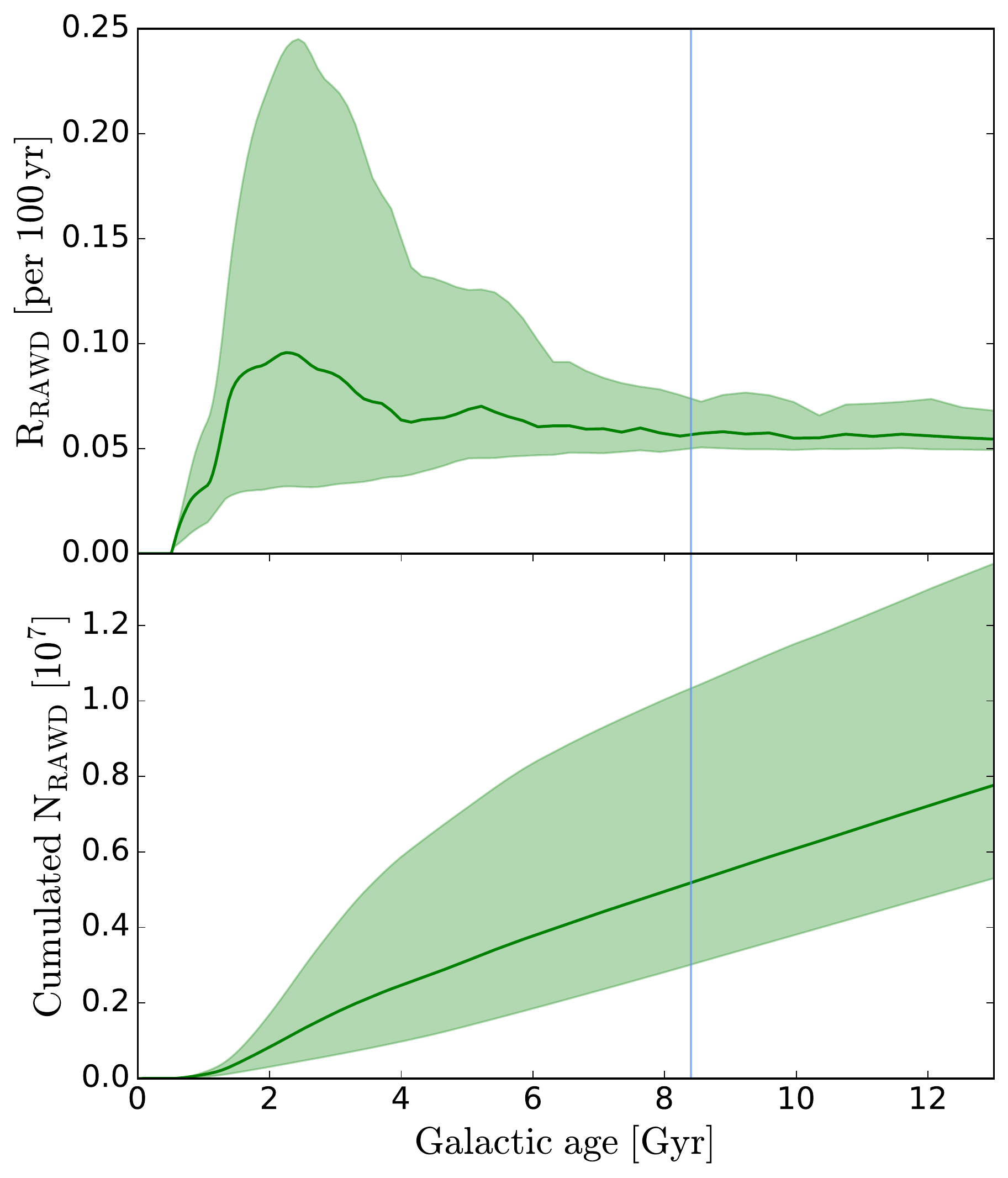}
\caption{Predicted rate (upper) and cumulated number (bottom) of rapidly accreting white dwarfs (RAWDs) as a function of Galactic age. The green solid lines represent our fiducial model while the green shaded areas show the range of solutions generated by different chemical evolution paths (see Figure~\ref{fig_age_met}).}
\label{fig_nb_rate}
\end{figure}

The upper panel of Figure~\ref{fig_nb_rate} shows the RAWD birth rates as a function of Galactic age.  The rates are most uncertain at early times and vary by an order of magnitude at 2.5\,Gyr. This peak of uncertainty is caused by the sharp transition at $Z_\mathrm{met}=0.001-0.002$ above which RAWD rates in SSPs drop by an order of magnitude (Figure~\ref{fig_rawd_rates}).
The time for the Galactic gas to reach this transition metallicity depends on the chosen chemical evolution path (Figure~\ref{fig_age_met}). When the metallicity of the gas evolves slowly, more low-metallicity SSPs will be formed, which will increase the RAWD rates. On the other hand, when the metallicity of the gas evolves rapidly, SSPs will be more metal-rich on average and RAWD formation will be somewhat suppressed (Figure~\ref{fig_rawd_rates}).

In all the considered chemical evolution paths, the sharp transition metallicity mentioned above is reached within the first Gyr of evolution (see Section~\ref{sect_MW_prop}), which is why the scatter in the Galactic rate decreases after 2.5\,Gyr. The level of scatter stays relatively constant beyond solar metallicity (blue vertical line in Figure~\ref{fig_nb_rate}) since we did not calculate yields and DTD functions for RAWDs at $Z_\mathrm{met}>0.014-0.02$. Because our target observable is the Sun, we did not need to follow the GCE calculation beyond the adopted solar value. 
When the metallicity of the gas reached solar, we simply applied the highest-metallicity yields and rate for all subsequent SSPs that formed at later times. Our predictions for the current Galactic RAWD rates are higher by a factor of $2-3$ compared the lower limit estimated by \cite{2017ApJ...834L..10D} that were based on population synthesis predictions (\citealt{2014MNRAS.445.1912C}) for the single-degenerate SN~Ia scenario (see also \citealt{2009ApJ...699.2026R}).

The lower panel of Figure~\ref{fig_nb_rate} shows the cumulated number of RAWDs in our simulation as a function of Galactic age. Because of the different chemical evolution paths assumed at early times, the predicted number of RAWDs that contribute to the solar composition varies by a factor of $\sim$\,3.5.

\begin{figure*}
\includegraphics[width=7in]{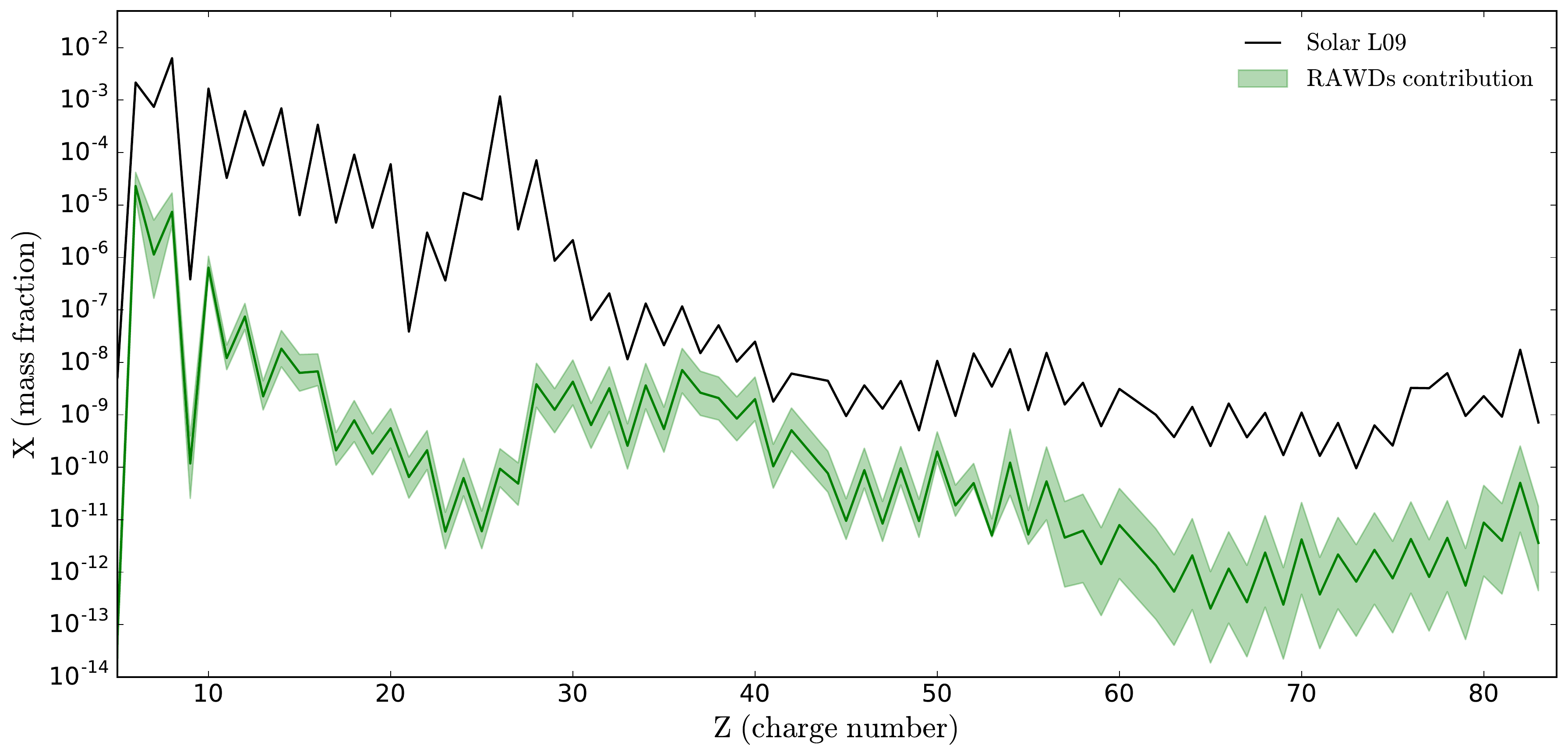}
\caption{Predicted contribution of rapidly accreting white dwarfs (green, RAWDs) to the elemental solar composition (black). The green solid line represents our fiducial model while the green shaded area shows the range of solutions generated by uncertainties in the total mass ejected by RAWDs (see Section~\ref{sect_rawd_impl}) and by different early chemical evolution paths (see Figure~\ref{fig_age_met}). The solar composition was taken from \protect\citet[L09]{2009LanB...4B...44L}.}
\label{fig_solar_full}
\end{figure*}

\begin{figure}
\center
\includegraphics[width=3.3in]{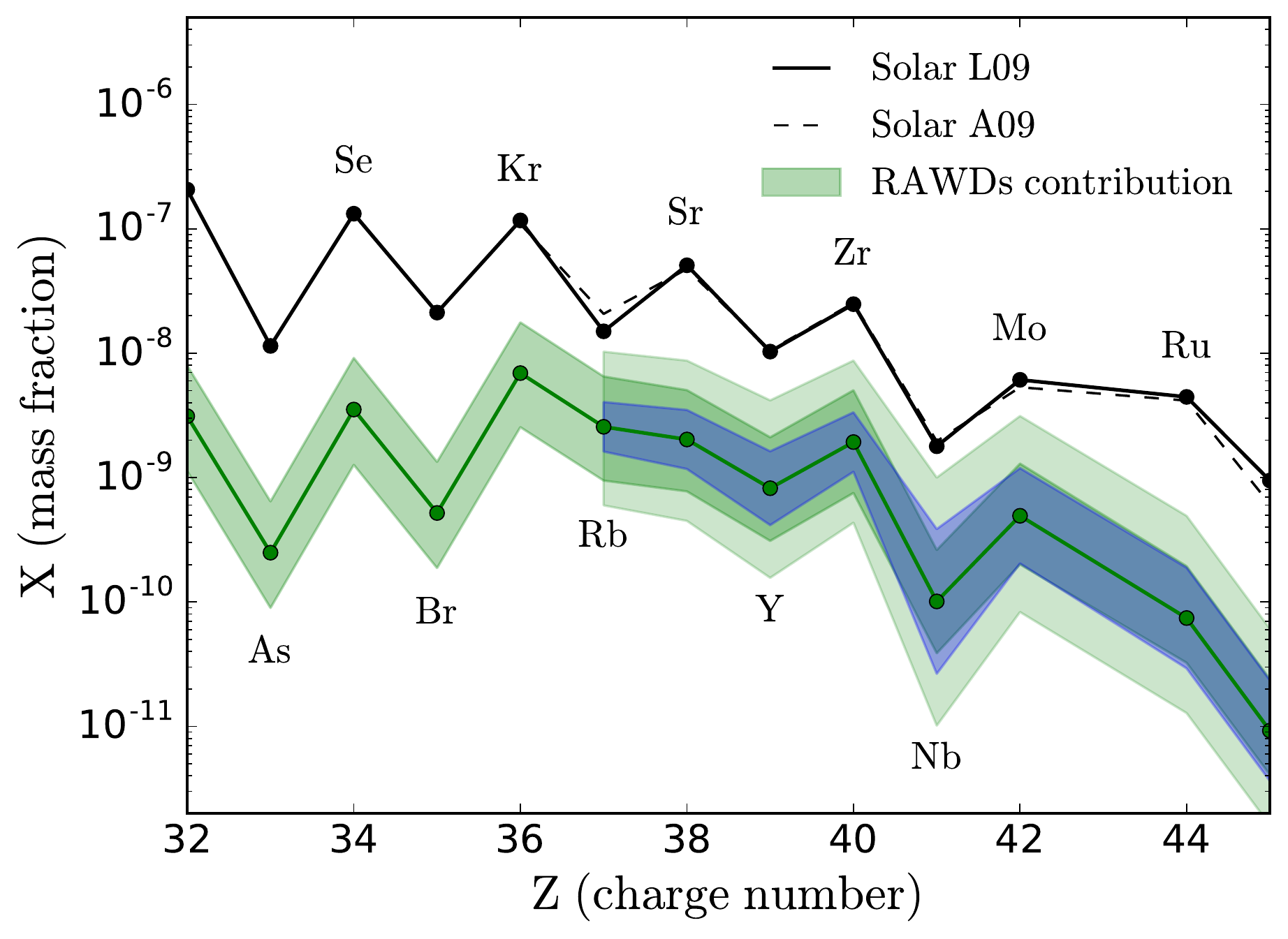}
\caption{Same as in Figure~\ref{fig_solar_full}, but zoomed on first-peak elements. The dashed black line shows the solar composition of \protect\citet[A09]{asplund09}.  The blue shaded area shows the uncertainties generated by nuclear reaction rates (see Section~\ref{sect_disc_y_uncert}). The larger lighter-green shaded area shows the combined uncertainties generated by different chemical evolution paths, different ejecta masses for each RAWD, and by nuclear reaction rate uncertainties.}
\label{fig_solar_zoom}
\end{figure}

\subsection{Elemental Composition} \label{sect_el_comp}
Figure~\ref{fig_solar_full} shows the predicted contribution of RAWDs to the chemical composition of the Galactic gas when the Sun forms, 4.6\,Gyr before the end of the simulation.  The green solid line represents our fiducial chemical evolution path (see dashed line in the bottom panel of Figure~\ref{fig_age_met}) assuming 0.75\,M$_\odot$ for the integrated \ipr ejecta over the lifetime of each RAWD. The green shaded area shows the range of solutions generated by using different chemical evolution paths (Figure~\ref{fig_age_met}) and different ejecta masses between 0.5 and 1\,M$_\odot$. This level of uncertainty varies from one element $Z$ to another because of the metallicity-dependent rates and yields adopted for RAWDs.

The level of uncertainty is systematically higher at $Z\gtrsim55$.  When the chemical evolution path favours low-metallicity SSPs ($Z_\mathrm{met}<0.002$), which occurs when the metallicity of the Galactic gas evolves slowly, there will be more RAWDs because of the higher birthrates predicted by our population synthesis model (Figure~\ref{fig_rawd_rates}). In addition, our RAWD yields at $Z_\mathrm{met}<0.002$ mainly produce elements with $Z\gtrsim55$ (Figure~\ref{fig_iRAWD_Y}).  The opposite situation occurs when the chemical evolution path favours high-metallicity SSPs ($Z_\mathrm{met}>0.002$). In that case, there will be less RAWDs along with a lack of nucleosynthetic production for $Z\gtrsim55$.
 
The situation is different for lighter elements (e.g, $Z=[30-55]$).  When low-metallicity SSPs are favoured, although more RAWDs will form compared to high-metallicity SSPs, less $Z\lesssim55$ elements will be ejected per RAWD event (Figure~\ref{fig_iRAWD_Y}). When high-metallicity SSPs are favoured, more $Z\lesssim55$ elements will be ejected per RAWD event, but less RAWDs will form in total.
To summarize, there is a cancelation effect in the mass of $Z\lesssim55$ elements ejected in the Galactic gas: high RAWD rates imply low nucleosynthetic yields, and vice-versa.  On the other hand, there is an amplification effect for the heavier $Z\gtrsim55$ elements (see paragraph above), which explains the larger spread seen for the heaviest elements in Figure~\ref{fig_solar_full}.

 Overall, the contribution of RAWDs to the solar composition is not significant except for elements near the first peak ($Z=[36-42]$).  Figure~\ref{fig_solar_zoom} shows a zoom of the region of interest.  According to our model, even though RAWDs are not the dominant contributor to the production of these elements, their contribution is still significant and could explain the origin of a fraction of the solar first-peak composition (see Table~\ref{tab_contribution_RAWDs}).

 \begin{deluxetable*}{ccccccc}
\tablewidth{0pc}
\tablecaption{Predicted contribution, in percentage, of rapidly accreting white dwarfs (RAWDs) to the first-peak elemental solar composition of \protect\citet[A09]{asplund09} and \protect\citet[L09]{2009LanB...4B...44L}.  The fiducial values represent the solid green line in Figure~\ref{fig_solar_zoom}. The values in bracket show the range of plausible solutions if we account for different galactic chemical evolution (GCE) paths, nuclear reaction rate uncertainties (Nucl. React.), and for both sources of uncertainties combined.  We described how we combined uncertainties at the end of Section~\ref{sect_el_comp}. \label{tab_contribution_RAWDs}}
\tablehead{ \colhead{\multirow{2}{*}{Z}} & \colhead{\multirow{2}{*}{Element}} & \multicolumn{5}{c}{RAWDs contribution [\%]} \\
\colhead{} & \colhead{} & \colhead{Fiducial (L09)} & \colhead{GCE (L09)} & \colhead{Nucl. React. (L09)} & \colhead{Combined (L09)} & \colhead{Combined (A09)} }
\startdata
35 & Br & 2.5 & [ 0.9 - 6.3 ] & --- & [ 0.9 - 6.3 ] & --- \\
36 & Kr & 5.9 & [ 2.2 - 15.2 ] & --- & [ 2.2 - 15.2 ] & [ 2.3 - 16.2 ] \\
37 & Rb & 17.1 & [ 6.3 - 43.4 ] & [ 10.8 - 27.0 ] & [ 4.0 - 68.6 ] & [ 2.9 - 49.7 ] \\
38 & Sr & 4.0 & [ 1.5 - 9.9 ] & [ 2.3 - 6.9 ] & [ 0.9 - 17.1 ] & [ 0.9 - 18.3 ] \\
39 & Y  & 8.0 & [ 3.0 - 20.5 ] & [ 4.0 - 15.8 ] & [ 1.5 - 40.5 ] & [ 1.5 - 39.6 ] \\
40 & Zr & 7.8 & [ 3.0 - 20.3 ] & [ 4.5 - 13.5 ] & [ 1.8 - 35.2 ] & [ 1.7 - 34.4 ] \\
41 & Nb & 5.7 & [ 2.2 - 14.7 ] & [ 1.5 - 21.6 ] & [ 0.6 - 56.0 ] & [ 0.5 - 51.0 ] \\
42 & Mo & 8.1 & [ 3.3 - 21.2 ] & [ 3.3 - 19.5 ] & [ 1.4 - 51.2 ] & [ 1.6 - 58.7 ] \\
44 & Ru & 1.7 & [ 0.7 - 4.4 ] & [ 0.7 - 4.2 ] & [ 0.3 - 11.1 ] & [ 0.3 - 11.8 ] \\
\enddata
\end{deluxetable*}

 \begin{figure*}
\includegraphics[width=7in]{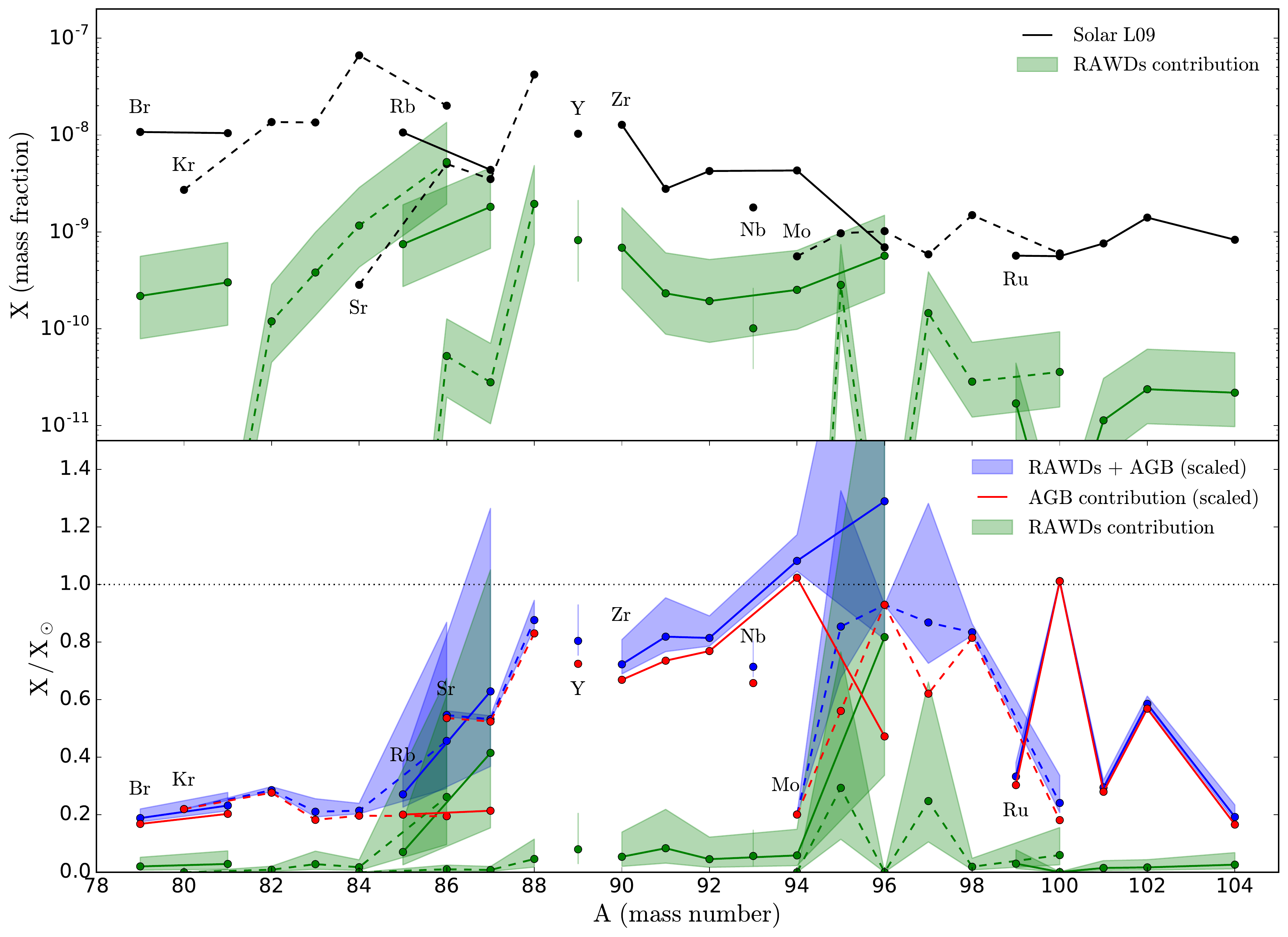}
\caption{Upper: Same as in Figure~\ref{fig_solar_zoom}, but decomposed in isotopic compositions.  The solar isotopic composition found in \protect\citet[L09]{2009LanB...4B...44L} are the same as in \protect\cite{asplund09}. The alternance between solid and dashed lines is to help keeping track of the isotopes with the same charge number. Lower: Contributions of AGB stars \protect\citet[$s$-process, red lines]{2015ApJS..219...40C}, RAWDs (green lines), and the sum of AGB and RAWDs (blue lines), in mass fraction relative to the solar composition (black dotted line). The AGB stars contribution has been reduced by 35\% so that the \sprn produces 100\% of $^{150}$Sm, an $s$-only isotope.}
\label{fig_solar_zoom_isol}
\end{figure*}

As shown in Figure~\ref{fig_solar_zoom} and Table~\ref{tab_contribution_RAWDs}, nuclear reaction rate and galactic chemical evolution uncertainties affect our results in a similar way.  To include nuclear reaction rate uncertainties in our fiducial prediction (blue shaded area in Figure~\ref{fig_solar_zoom}), we used the 1-$\sigma$ dispersions extracted from normal distributions generated by Monte Carlo calculations (see Section~\ref{sect_disc_y_uncert}).  The same dispersions have been applied to our lower and upper limit predictions, which were produced by assuming different chemical evolution paths and ejecta masses, in order to estimate the combined uncertainties (larger and lighter green shaded area).
\\
\\
\subsection{Isotopic Composition} \label{sect_iso}
The upper panel of Figure~\ref{fig_solar_zoom_isol} shows the contribution RAWDs to the isotopic composition of the solar composition for the same elements shown in Figure~\ref{fig_solar_zoom}.  Our predictions do not overproduce any isotope, except for $^{96}$Zr which is produced in a similar quantity than what is observed in the Sun.  In general, the isotope production patterns of our \ipr yields do not follow the solar composition.

In the bottom panel of Figure~\ref{fig_solar_zoom_isol}, we divided our predictions by the solar composition and compare the RAWDs contribution with the \spr isotopic pattern predicted by our fiducial GCE model using the non-rotating AGB stars yields from the FRUITY database (\citealt{2015ApJS..219...40C}). We scaled down the \sprn by 35\,\% so that it accounts for 100\,\% of the $^{150}$Sm observed in the Sun, which is an $s$-only isotope. This normalization is consistent with \cite{2015ApJ...801...53C} who also noticed an overestimation of about 45\,\% for $s$-only isotopes using their non-rotating AGB yields, but using a different GCE code.  Using their rotating AGB models would likely underestimate $^{150}$Sm (see their Figure~6).  We do not include the isotopic composition of the \rprn because of the large uncertainties associated with theoretical calculations (e.g., \citealt{2016PhRvL.116l1101M,2016PrPNP..86...86M}).  Using the \rpr residuals as an alternative solution would leave, by definition, no room for the $i$~process.

The blue lines represent the combined contribution of RAWDs and AGB stars. Uncertainties in the yields of AGB stars is not included in this panel.
In some cases, as shown in the bottom panel of Figure~\ref{fig_solar_zoom_isol}, the \iprn (green lines) has a production peak where the \sprn (red lines) has a local minima (e.g., $^{96}$Zr, $^{97}$Mo).  
In the case of $^{96}$Mo, the \iprn shows a local minima while the \sprn shows a global maxima. Although isotope yields for RAWD and AGB models need to be addressed with quantified uncertainties, which is beyond the scope of this paper, Figure~\ref{fig_solar_zoom_isol} suggests that the \iprn can complement the $s$~process for some isotopes. 

As an example, $^{95}$Zr represents a branching point (e.g., see \citealt{2014ApJ...780...95L,2016ApJ...827...30B}) which means that there is a probability to capture a neutron and form the stable $^{96}$Zr isotope, depending on the neutron density.  During the $i$ process, the neutron density is higher than with the \sprn and unstable $^{95}$Zr isotopes are more efficiently transformed into $^{96}$Zr, which leads to a higher $^{96}$Zr abundance compared to the \spr case.

\section{Discussion} \label{sect_disc}
Here we discuss the various sources of uncertainties unaccounted in our results and the limitations of
our galactic chemical evolution code to quantify the contribution of RAWDs to the solar composition.
 We also discuss the implications of our results on the solar LEPP and on the single-degenerate SN~Ia scenario.

\subsection{Common Envelope Evolution} \label{sect_disc_pop}
In the adopted (energy balance) common envelope formalism (see Section~\ref{sect_pop_synth}), the $\alpha_{\rm CE}$ and $\lambda$ parameters contain a lot of ``unknown physics" and the assumptions made during this phase are one of the largest sources of uncertainty in our models \citep[see][]{2013A&ARv..21...59I,2014A&A...562A..14T}. 
Higher values of $\alpha_{\rm CE} \times \lambda$ means higher ejection efficiencies which leads to wider orbital separations following the ejection of the CE. 
In general, choosing different reasonable values for these quantities could affect our results, but not in a drastic way. For example, if the physical processes leading to the unbinding of the CE were less efficient (e.g. lower $\alpha_{\rm CE}$ or $\lambda$ values), some fraction of the ``standard" binaries would not make RAWDs, as they would follow a different evolution that may cause them to merge too early on. However, binaries that would not have become RAWDs in our standard model, since they were not brought close enough together after the CE, would likely populate this RAWD parameter space instead.

\subsection{Effect of Metallicity on RAWD Birthrates} \label{sect_effect_Z_RAWDs}
We have shown that at higher ($\sim$solar) metallicities, the RAWD birthrate is about 10 times lower than for lower metallicities (Figure~\ref{fig_rawd_rates}). As described below, this is due to a combination of effects, which include metallicity-dependent stellar winds and different mass ratios for the stars when the companion transfers hydrogen towards the WD. 

One side effect of metallicity-dependent wind mass loss is that the lower-metallicity WDs will be more massive than their higher-metallicity counterparts, since the star was able to maintain larger (core) mass during later stages of stellar evolution. As a consequence, at time of Roche-Lobe overflow (RLOF) between the H-rich (e.g. Hertzsprung Gap) companion and the CO WD, {\em lower metallicity systems have less extreme mass ratios}. The less extreme mass ratio between the WD and the H-burning star is what enables these systems to undergo (quasi) stable mass transfer, and thus evolve into RAWD binaries. On the other hand, the higher-metallicity binaries have more extreme mass ratios at time of RLOF, which makes it more likely for them to encounter mass transfer on a dynamical timescale (e.g. CE evolution).

As noted in Section~\ref{sect_pop_synth}, there is a rather extreme decrease in RAWD birthrate between $Z_\mathrm{met}=0.001$ and 0.002.
Many systems which would make RAWDs involving evolved companions in the lower $Z_\mathrm{met}$ (0.0001, 0.001) models will make detached double WDs in the higher $Z_\mathrm{met}$ (0.002, 0.02) models consisting of a CO WD and a helium WD (due to stripping of the H-envelope during CE evolution).
The reason why the rate difference is notably extreme between our $Z_\mathrm{met}=0.001$ and $Z_\mathrm{met}=0.002$ models is because 
of a transition region within our \startrack\ algorithm that is used to determine whether a system in RLOF will encounter mass transfer on a thermal or dynamical timescale \citep[][see section 5]{2008ApJS..174..223B}.

We note that our algorithm that calculates the stability of mass transfer is uncertain, though star systems have indeed been observed to undergo a phase of quasi-stable mass transfer prior the (expected) CE phase, at least in massive stars. If this same analogy can be applied to lower-mass stars, we may be underestimating the RAWD birthrates at high metallicities, in which case our imposed criteria for undergoing a CE should be revised to allow the production of more RAWDs at more extreme stellar mass ratios.  In Annexe~\ref{annexe_A}, we explore an alternative set of GCE predictions where the RAWD birthrate smoothly evolves as a function of metallicity without a sharp transition.

There is another reason why higher-metallicity systems do not produce as many RAWDs, which is applicable to a different evolutionary channel (WD+MS RAWD): some higher-metallicity binaries are more readily destroyed via mergers during the first mass transfer event when the primary is on the RGB and the secondary is still on the MS. For the lower-metallicity counterpart, the (smaller in radius) primary would have already reached the early AGB and thus would have a larger core mass than the higher-metallicity RGB primary counterpart, despite the lower wind-mass loss rates in the lower-metallicity model. Both star systems will go to CE, but only the low-metallicity system will survive, leaving behind a He-burning sub-giant and a MS star, which eventually evolves into a RAWD. The higher-metallicity system that has the more extreme mass ratio (between core and MS star) will end up as a merger between a compact He-burning core and a MS star. 

\subsection{Implications of Mass Retention Efficiency on the Type Ia Supernova Rate} \label{sect_sds}
One of the leading progenitor scenarios of SNe~Ia includes the `textbook' single degenerate (SD) scenario, in which a CO WD approaches the Chandrasekhar mass limit via accretion from a (usually hydrogen-rich) stellar companion. Studies tracking the theoretical evolution of interacting binary populations have shown that it is difficult for  a CO WD to build up to the Chandrasekhar mass via hydrogen accretion \citep[e.g.,][but see \citealt{2004MNRAS.350.1301H}]{2009ApJ...699.2026R,2013A&A...552A..24B}. In addition, recent works have shown that some, if not most, SN~Ia explosions may be more easily explained by exploding sub-Chandrasekhar mass WDs, either via mergers, or `classic' double-detonations \citep[][see also \citealt{2016MNRAS.457.3254M}]{2012ApJ...747L..10P,2013ApJ...770L..35S,2016ApJ...827..128V,2016ApJ...821...67S,2017ApJ...841...48S}. Despite the recent favouritism for sub-Chandrasekhar mass models (see also \citealt{2017arXiv171005030M}), the Chandrasekhar mass SD scenario \citep[sometimes referred to as the delayed detonation scenario][]{2013A&A...559A.117C} still remains a viable progenitor candidate \citep[e.g.,][]{2012ApJ...758..123W,2013A&A...559L...5S,2015ApJ...805..150F,2015ApJ...801L..31Y,hitomi17}.  However, the measurement of nebular emission lines in different galaxies implies a limit on the contribution of the SD scenario to the overall observed SNe~Ia rate to less than $\sim10$\,\% (\citealt{2014MNRAS.442.1079J,2016MNRAS.461.4505J}, see also \citealt{2017NatAs...1E.263W} and \citealt{2018ApJ...852L...6B}).

In our adopted accretion model for RAWDs, where accretion on CO WDs is suppressed at relatively high mass transfer rates (see Section~\ref{sect_pop_synth}), it would be (nearly) impossible to produce any SNe~Ia via the `textbook' SD channel where a hydrogen-rich donor transfers mass via stable RLOF. We do find however a relatively small number of CO WDs that accrete up to the Chandrasekhar mass via wind accretion when the donor is an AGB star. These systems undergo a different evolutionary channel from RAWDs, where RLOF phases occur between an evolved star and a MS companion, so they never enter the RAWD parameter space.

When the primary star turns into a CO WD, it is already fairly close to the Chandrasekhar mass (${\sim} 1.38$ \msun), and futher accretion by the AGB companion wind is able to push the WD toward the Chandrasekhar mass. We predict that these SN~Ia progenitors, if realised in nature, have prompt delay times ($< 100$ Myr), and only occur in higher metallicity populations ($Z_{\rm met} \geq 0.002$). The lower metallicity primary stars (which experience less wind mass loss) are more likely to evolve into ONe WDs rather than CO WDs. 
We note that our current study cannot rule out near-Chandrasekhar mass explosions via RLOF from helium-rich companions, which have been proposed as good candidates for thermonuclear supernovae, in particular the fainter SN~Iax-likes \citep{2015MNRAS.450.3045K,2015A&A...573A...2S}

\subsection{Yield Uncertainties} \label{sect_disc_y_uncert}
The \ipr yields predicted with the RAWD models (Figure~\ref{fig_iRAWD_Y}) depend on stellar physics and nuclear reaction
rate uncertainties that are translated into yield uncertainties. Various stellar physics uncertainties will be analyzed
elsewhere, and in that paper we will provide a detailed discussion of our RAWD models. Here, we only report some results on the yield 
uncertainties that are linked to the uncertainties of the (n,$\gamma$) cross sections of unstable
isotopes near the magic neutron number $N=50$ and that are relevant to our predicted contribution of RAWDs to the solar first-peak
elemental abundances (all details of the corresponding uncertainty study are presented in \citealt{2016arXiv161101121D}). 
These results have been obtained for a model of Sakurai's object, whose \ipr yields
are similar to those of RAWD models with a nearly solar metallicity \citep[Figure~4 in][]{2017ApJ...834L..10D}. 
The following paragraphs give a brief description of what has been done.

\begin{figure}
\center
\includegraphics[width=3.4in]{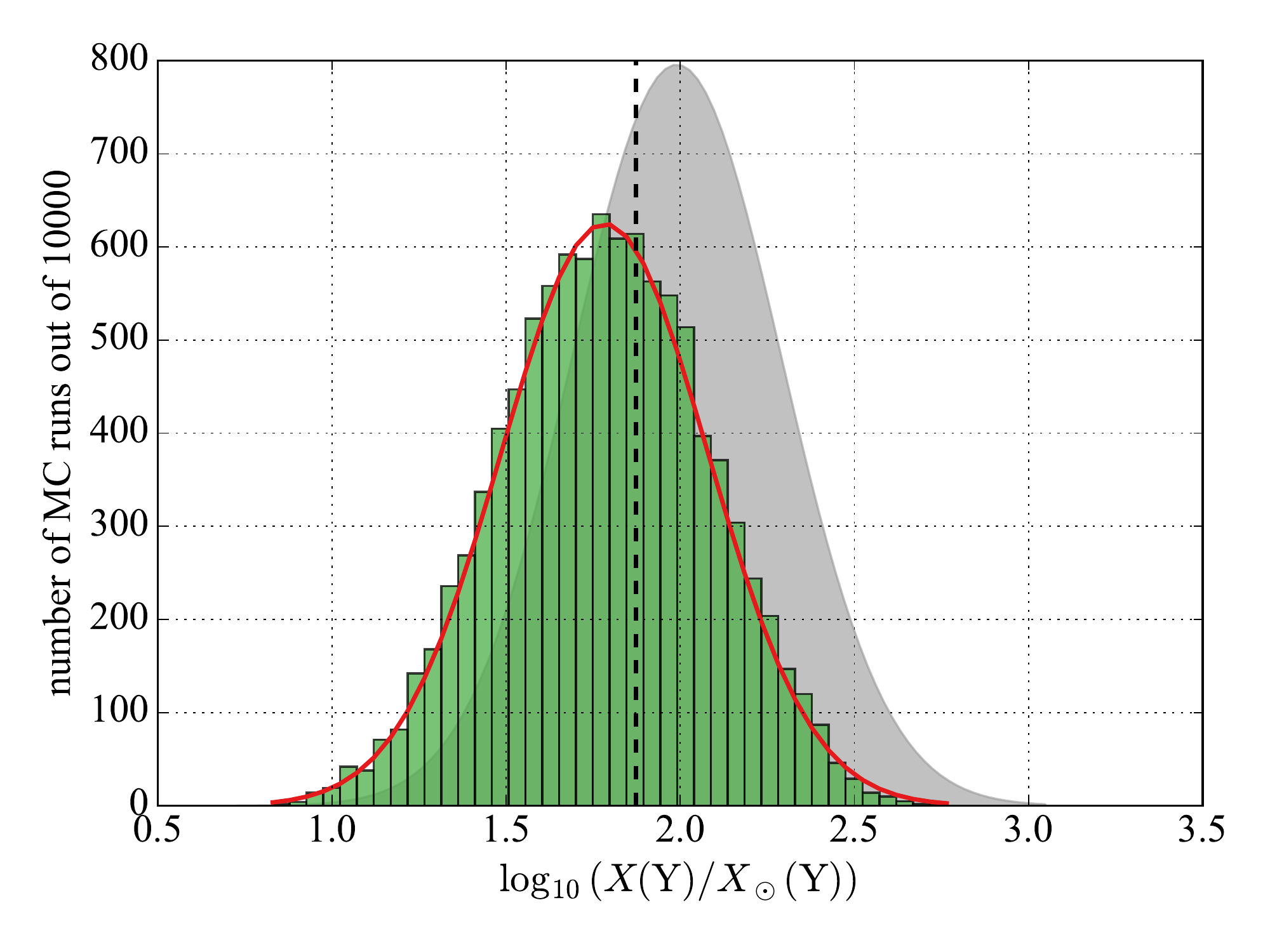}
\caption{A distribution of the predicted abundance of Y (mass fraction) from our MC simulations in which we randomly varied the (n,$\gamma$)
    cross sections of 52 unstable isotopes near $N=50$ (the green histogram).
    The vertical dashed line is the benchmark model prediction for the reaction-rate multiplication factors $f_i = 1$.
    By fitting a normal distribution (the red curve),
    we estimate the mean and dispersion of the results. For comparison, the grey-shaded area shows the observed Y abundance
    represented by a normal distribution with the mean and dispersion from \cite{1999A&A...343..507A}.}
\label{fig_Y_hist}
\end{figure}

First, 52 unstable isotopes of Br, Kr, Rb, Sr, Y, and Zr, whose (n,$\gamma$) cross sections can potentially affect the predicted
abundances of Rb, Sr, Y and Zr, that were also measured in Sakurai's object by \cite{1999A&A...343..507A}, have been selected
from the chart of nuclides. Because there is no experimental information on the (n,$\gamma$) cross sections of these isotopes,
the Hauser-Feshbach model of a statistical decay of a compound nucleus was used to obtain it. When systematically
varying between available five nuclear level density models and four $\gamma$ ray strength function parametrizations within the Hauser-Feshbach
code, the largest and smallest n-capture rates were found and their ratios were assigned to 
the maximum variation factors $v_i^\mathrm{max}$ for all of the 52 isotopes. 

Second, we have carried out Monte-Carlo (MC)
simulations in which the \ipr nucleosynthesis in Sakurai's object was modeled 10{,}000 times with randomly selected
sets of multiplication factors $f_i$ for the (n,$\gamma$) reaction rates involving the 52 selected isotopes
(the benchmark model, with which we compared the results of our MC simulations, had all $f_i = 1$).
For each of the 52 isotopes in each of the 10{,}000 MC runs, we first selected a random number $v_i^\mathrm{rand}$ from
a uniform distribution in the interval $1\leq v_i^\mathrm{rand} \leq v_i^\mathrm{max}$, then the multiplication factor
was assigned a value of $f_i = (p/v_i^\mathrm{rand}) + (1-p)v_i^\mathrm{rand}$, where $p$ were assumed to take a value of 0
or 1 with an equal probability.  

The green histogram in Figure~\ref{fig_Y_hist} shows a distribution of the predicted abundance of Y from our MC simulations.
Similar histograms were constructed for the other elements of the fist peak. By fitting them with normal distributions,
we were able to estimate their mean values and dispersions. For Rb, Sr, Y, Zr, Nb, Mo, and Ru, the (n,$\gamma$) reaction rate
uncertainties of the 52 unstable isotopes are translated into the predicted yield uncertainties of 0.20, 0.24, 0.30,
0.24, 0.58, 0.38, and 0.40 (for the distributions of the logarithmic abundance ratios with respect to the initial or
solar abundances). For the first four elements,
these uncertainties turn out to be less than or comparable to their observed errors from \cite{1999A&A...343..507A}.
For the rest three elements, the estimated uncertainties do not include a contribution from the (n,$\gamma$) reaction
rate uncertainties of unstable isotopes heavier than Zr, therefore they can in fact be (probably, slightly) different.
Nevertheless, we used all of these data in the analysis of our predicted contribution of RAWDs to the solar first-peak elemental abundances.

\subsection{Galaxy Evolution Uncertainties}
Because the yields and rates used for RAWDs are metallicity-dependent, our results 
are affected by the chemical evolution path used in our Milky Way model.
The overall evolution of metallicity as a function of time in a one-zone model
is driven by the shape of the star formation history (SFH) and by the
amount of gas in which stellar ejecta are deposited. The latter is 
controlled by the star formation efficiency. As shown in Figure~\ref{fig_age_met},
it is possible to create different chemical evolution paths using
the same SFH but by varying the star formation efficiency.
But the shape of the SFH also plays an important role in GCE, as it defines
how many SSPs are formed at a specific metallicity and how fast the galactic
gas is being enriched.  Indeed, as shown in \cite{2003PASA...20..189F}, different SFHs can also lead to different
chemical evolution paths at early times (see also \citealt{2017MNRAS.466.2474H}).

In this work, we only varied the star formation efficiency, but any variation from what we 
assumed for the SFH could change the predicted number ratio
of low- to high-metallicity SSPs and thus affect our predictions.  This source of uncertainty has also been
discussed by \cite{2015ApJ...801...53C} in the context of metallicity-dependent AGB star yields.
As explained in Sections~\ref{sect_rawd_rates} and \ref{sect_el_comp}, our results
are sensitive to the age-metallicity relationship, and thus the SFH, during the first Gyr of evolution.
Within a cosmological context, the SFH of galaxies in the early universe is significantly
affected by structure formation and by galaxy mergers (e.g., \citealt{2012ApJ...745...50W}). 
The stochastic early phase of the Milky Way is still not well constrained, and our one-zone
model is not suited to address this complexity (but see \citealt{2017arXiv171006442C}).

In addition, the concept of a direct correlation between age and metallicity breaks
down at very low metallicity. Hydrodynamic simulations have shown that non-uniform
mixing of stellar ejecta at early times generate
significant scatter in the age-metallicity space (e.g., \citealt{2012ApJ...745...50W,2015ApJ...814...41H,2017MNRAS.465.2212S}).
Accounting for more metallicity dispersion in our model would modify the metallicity
range associated with our SSPs, which could affect the predicted contribution of
RAWDs, given their strong dependency on metallicity.  However, it is difficult
with our model to evaluate whether those non-uniformities would significantly
affect our results compared to an averaged uniformly-mixed model.

By using a one-zone model, we do not account for the formation
timescale of different Galactic components such as the halo, the
thick disc, and the thin disc.  With multi-zone models (e.g.,
\citealt{1992ApJ...387..138F,1995ApJ...444..207P,1999ApJ...521..691T,2014ApJ...787...10B}), the formation of the Galactic disc is
delayed relative to the formation of the halo.  
Assuming our one-zone model represents the Galactic disc, the time at which 
we form the Sun could be reduced by $\sim$\,1\,Gyr, which is the
typical delay for disc formation (\citealt{1995ApJ...444..207P,2001ApJ...554.1044C}). According to
 the bottom panel of Figure~\ref{fig_nb_rate}, this formation delay
would change the total number of RAWDs included in the solar
composition by no more than 20\%, assuming the same star
formation history.

\subsection{Solar Lighter Element Primary Process}
By combining the \rprn and the weak and main $s$ processes in a galactic chemical
evolution context, the solar composition near the first peak up to Xe is not fully explained without introducing an additional 
lighter element primary process, the so-called LEPP (\citealt{2004ApJ...601..864T,2007ApJ...671.1685M}).  This claim has later on been confirmed
by \cite{2014ApJ...787...10B}.  According to \cite{2004ApJ...601..864T}, the unaccounted fractions are
8\,\% for Sr, 18\,\% for Y, Zr, Nb, and 25\,\% for Mo (see their Table~4).
As shown in our Table~\ref{tab_contribution_RAWDs}, RAWDs could explain
the LEPP for some elements. The production of Sr in RAWDs
is about half the production of Y and Zr, a specific feature associated
with the LEPP. For elements heavier than Mo ($Z>42$), the contribution of RAWDs
drops and becomes insignificant (see Figure~\ref{fig_solar_full}).

However, the need for the solar LEPP is still a matter of debate.  \cite{2013ApJ...762...31P} investigated
the impact of uncertainties in the $^{12}$C\,+\,$^{12}$C reaction rate and
found that massive stars could produce enough first-peak elements to fill
the missing LEPP.  In addition, \cite{2015ApJ...801...53C} showed that the need for the LEPP
depends on the physics involved in modeling AGB stars and on the star
formation history adopted in GCE models.  It is beyond the scope of this
paper to address the solar LEPP in more details. But our results suggest
that RAWDs could provide an important fraction of the solar composition for Sr, Y, and Zr. We also need to keep in mind that in the calculations of \cite{2014ApJ...787...10B}, the s-only isotopes were also missing in relevant amounts, but s-only isotopes are not made efficiently in RAWDs (see for instance the case of the s-only isotope $^{96}$Mo discussed in Section~\ref{sect_iso}).

\section{Conclusion} \label{sect_conc}
We introduced RAWDs, which stands for rapidly accreting white dwarfs, in our NuPyCEE
framework to quantify in a GCE context their contribution to the solar
composition. To do so, we calculated metallicity-dependent \ipr yields using
\mesa\ and \mppnp\ (Figure~\ref{fig_iRAWD_Y}) and delay-time distribution functions using \startrack\ 
(Figure~\ref{fig_rawd_rates}), and applied them to all stellar populations
formed in our Milky Way model.  We tested different normalizations for the
mass ejected by RAWDs and different chemical evolution paths to reach
solar metalllicity by the time the Sun forms  (Figure~\ref{fig_age_met}).

Our yields and rates for RAWDs are very sensitive to metallicity. Yields at $Z_\mathrm{met}=0.014$
produce roughly 3 orders of magnitude more Sr, Y, and Zr than yields at
$Z_\mathrm{met}=0.00014$ (low-metallicity yields tends to produce heavier
elements). Rates at low metallicity are higher by an order of magnitude compared to
the ones at high metallicity, with a sharp transition occurring between $Z_\mathrm{met}=0.001$
and 0.002 (but see Section~\ref{sect_effect_Z_RAWDs}).  Because of these dependencies, the impact of the chemical evolution
path on the predicted contribution of RAWDs varies from one element to another,
with the heaviest elements ($Z\gtrsim$\,55) being the most uncertain (Figure~\ref{fig_solar_full}).

We found that RAWDs can have a significant contribution to the solar
composition for elements near the first \spr peak: $[2-15]$\,\% for Kr, [$6-43$]\,\% for Rb,
[$2-10$]\,\% for Sr, [$3-21$]\,\% for Y, [$3-20$]\,\% for Zr, [$2-15$]\,\% for Nb,
and [$3-21$]\,\% for Mo.  Uncertainties associated with population synthesis models are discussed in Section~\ref{sect_disc}.  When nuclear reaction rate uncertainties for the \iprn are included in our GCE predictions,
the upper boundaries increase by a factor of $1.5-2$ for Rb, Sr, Y, and Zr,
and by a factor of 3.8 and 2.4 for Nb and Mo, respectively (see Table~\ref{tab_contribution_RAWDs}).
This highlights the importance of creating and maintaining communication between
nuclear astrophysics and galaxy evolution, as both fields can have a significant
impact on the predicted evolution of chemical elements using galaxy models.
 
We found that the \iprn could complement the \sprn in reproducing the solar composition
for some isotopes (e.g., $^{96}$Zr, $^{95}$Mo, and $^{97}$Mo).  Given the uncertainties in our predictions,
our work shows that RAWDs could explain a fraction of the solar LEPP, especially
for Sr, Y, and Zr.  Within the limitations of our models (see Section~\ref{sect_disc}), we confirm the  
calculation made by \cite{2017ApJ...834L..10D} showing that RAWDs are relevant to
the chemical evolution of first-peak elements.  We predict a current
Galactic RAWD rate of about $5\times10^{-4}$~yr$^{-1}$.

Observationally, RAWD systems should appear as super-soft X-ray sources most of the time, unless being (easily) obscured 
by interstellar or circum-binary matter \citep{vandenheuvel:92}. The latter factor (see also \citealt{woods:16}) may explain 
why only a few RAWD candidates out of theoretically predicted dozens were found in the Large and Small Magellanic Clouds \citep{lepo:13}.

Our work illustrates the contribution of \ipr nucleosynthesis on the solar composition
and is thus complementary to previous studies that discussed the presence of \ipr
signatures in metal-poor stars (\citealt{2016ApJ...831..171H,2016ApJ...821...37R,2017arXiv171001763C}).

\acknowledgments
We are thankful to the anonymous referee and to Maria Lugaro for useful discussions.
This research is supported by the National
Science Foundation (USA) under Grant No. PHY-1430152 (JINA Center
for the Evolution of the Elements).
The collaboration uses services of the Canadian Advanced Network for Astronomy Research (CANFAR) which in turn is supported by CANARIE, Compute Canada, University of Victoria, the National Research Council of Canada, and the Canadian Space Agency. BC acknowledges support from the ERC Consolidator Grant (Hungary) funding scheme (project RADIOSTAR, G.A. n. 724560).
AJR is supported by the Australian Research Council Centre of
Excellence for All-sky Astrophysics (CAASTRO) through project number
CE110001020. KB acknowledges support from the Polish National Science Center (NCN) grant: Sonata Bis 2 (DEC-2012/07/E/ST9/01360).

\software{{\tt OMEGA} \citep{2017ApJ...835..128C},
                {\tt SYGMA} \citep{ritter_sygma},
                {\tt NuPyCEE} (\url{http://nugrid.github.io/NuPyCEE}),
                {\tt MESA} \citep{2013ApJS..208....4P},
                {\tt mppnp} \citep{2016ApJS..225...24P},
                {\tt StarTrack} \citep{2002ApJ...572..407B,2008ApJS..174..223B},
                {\tt NumPy} \citep{2011arXiv1102.1523V},
                {\tt matplotlib} (\url{https://matplotlib.org}).}
                
\appendix
\section{Alternative RAWD Rates} \label{annexe_A}
\begin{figure}[h]
\center
\includegraphics[width=3.3in]{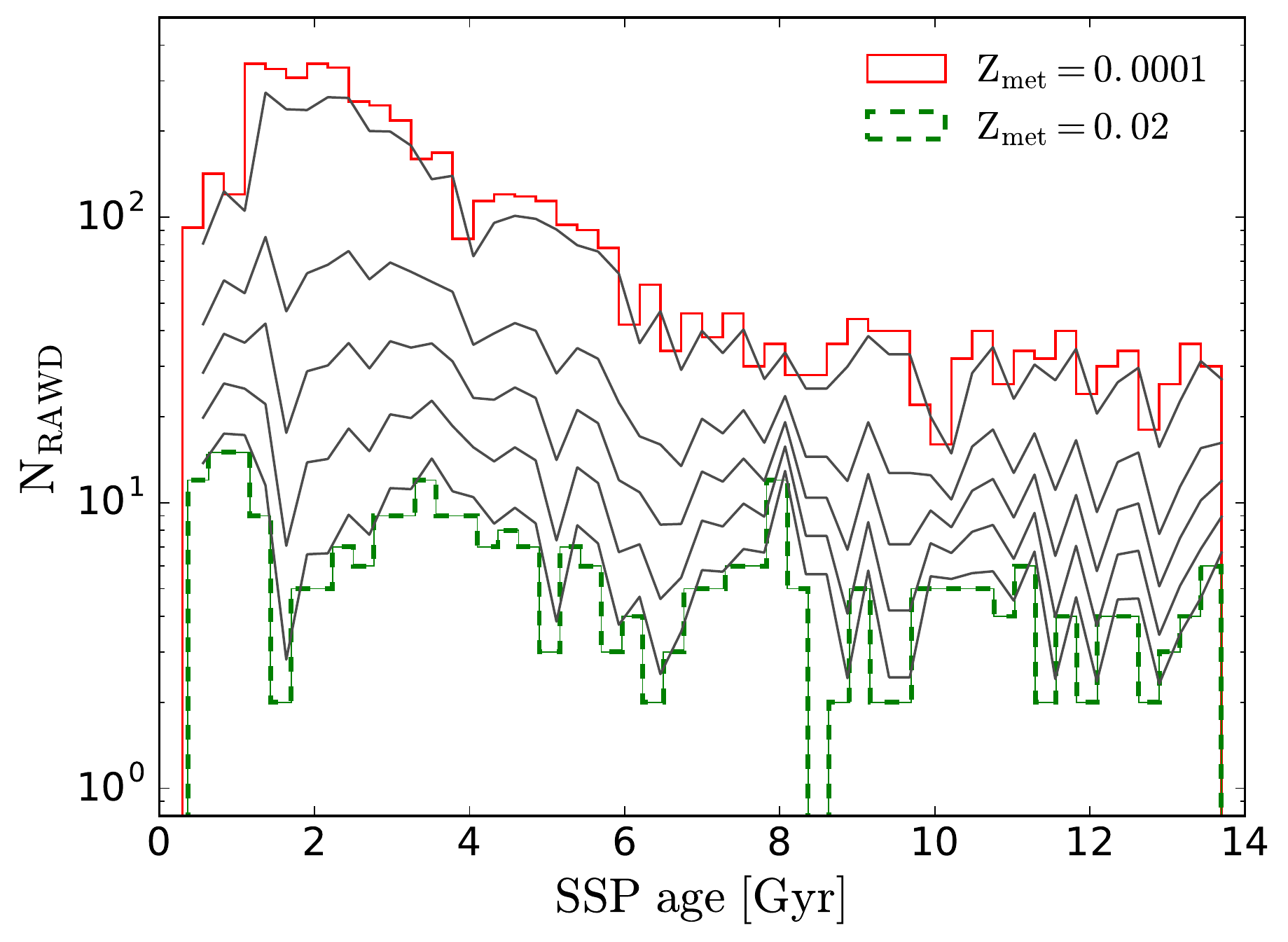}
\caption{Same as in Figure~\ref{fig_rawd_rates} but only accounting for $Z=0.0001$ and 0.02.  The grey lines are the linearly interpolated rates for the same metallicities as in RAWD yields (see Section~\ref{sect_yields}).  The interpolation has been made in the log-log space.}
\label{fig_rawd_rates_2}
\end{figure}

\begin{figure}
\center
\includegraphics[width=3.3in]{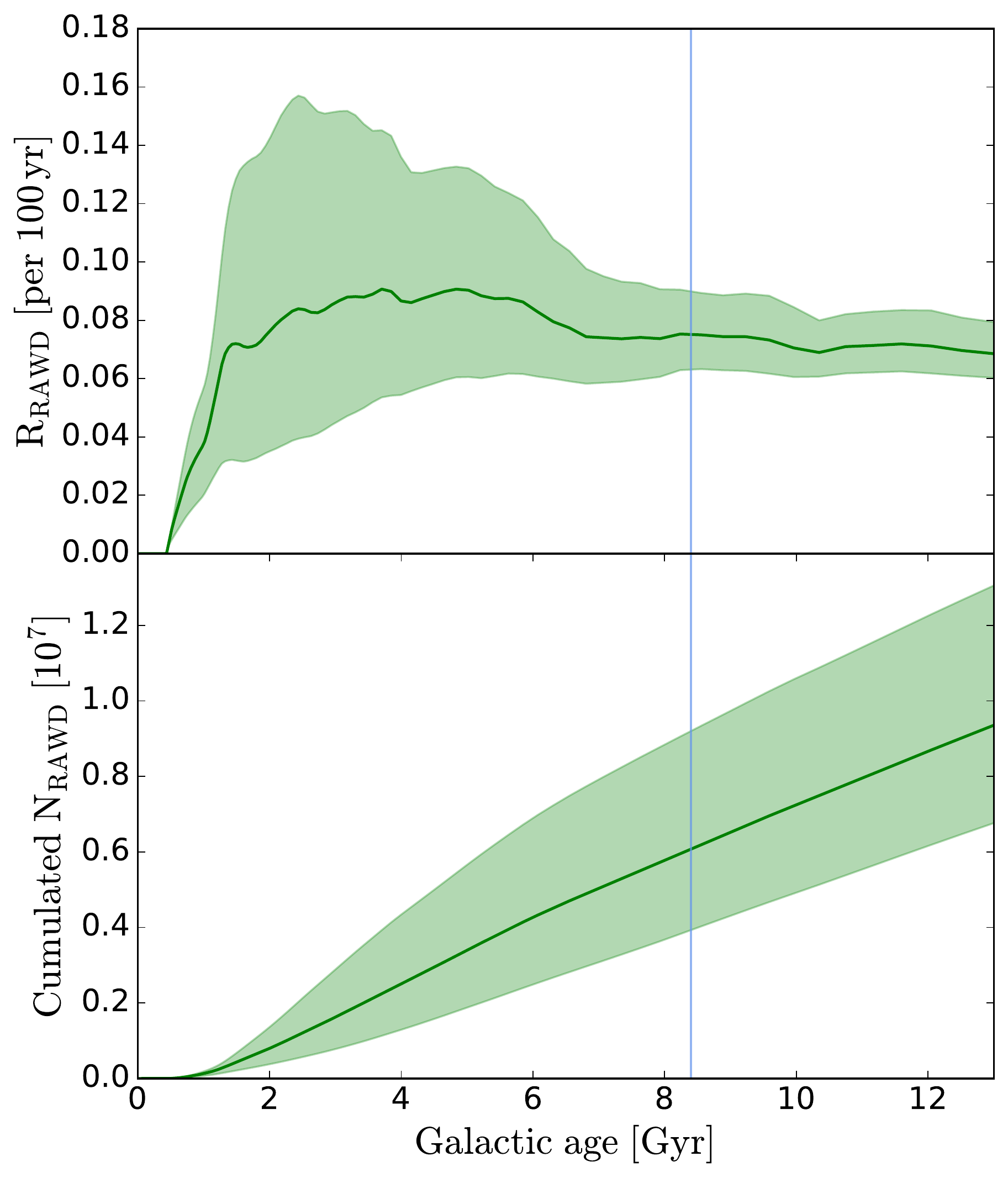}
\caption{Same as in Figure~\ref{fig_nb_rate}, but using the RAWD rates shown in Figure~\ref{fig_rawd_rates_2}.}
\label{fig_nb_rate_2}
\end{figure}

In Section~\ref{sect_effect_Z_RAWDs}, we discussed the impact of metallicity on the predicted RAWD rates as well as the origin of the sharp transition seen between $Z=0.001$ and 0.002 (see Figure~\ref{fig_rawd_rates}). Although this transition cannot be ruled out at the moment, it is still possible that the transition might be smoother.  More investigation is needed.  In order to test the sensitivity of our results on this sharp transition, we repeat in this section our calculations by linearly interpolating the predicted RAWD rates using the two extreme metallicities only ($Z=0.0001$ and 0.02).  This provides a smooth transition across all metallicities, as shown in Figure~\ref{fig_rawd_rates_2}.  Using this approach, the predicted contribution of RAWDs to the solar composition and the current Galactic rate are increased by about 25-40\% (see Figure~\ref{fig_nb_rate_2} and Table~\ref{tab_contribution_RAWDs_2}).  Although our results are sensitive to the metallicity-dependent RAWD rates, the magnitude of our predictions is not significantly affected by the sharp transition in the RAWD rate seen between $Z=0.001$ and 0.002.

\begin{deluxetable}{ccccccc}
\tablewidth{0pc}
\tablecaption{Same as in Table~\ref{tab_contribution_RAWDs}, but using the RAWD rates shown in Figure~\ref{fig_rawd_rates_2}.
\label{tab_contribution_RAWDs_2}}
\tablehead{ \colhead{\multirow{2}{*}{Z}} & \colhead{\multirow{2}{*}{Element}} & \multicolumn{5}{c}{RAWDs contribution [\%]} \\
\colhead{} & \colhead{} & \colhead{Fiducial (L09)} & \colhead{GCE (L09)} & \colhead{Nucl. React. (L09)} & \colhead{Combined (L09)} & \colhead{Combined (A09)} }
\startdata
35 & Br & 3.4 & [ 1.3 - 8.1] & --- & [ 1.3 - 8.1] & --- \\
36 & Kr & 8.3 & [ 3.3 - 19.5 ] & --- & [ 3.3 - 19.5 ] & [ 3.5 - 20.9 ]\\
37 & Rb & 24.0 & [ 9.5 - 55.8 ] & [ 15.2 - 38.0 ] & [ 6.0 - 88.3 ] & [ 4.3 - 63.9 ]\\
38 & Sr & 5.6 & [ 2.3 - 12.8 ] & [ 3.26 - 9.7 ]  & [ 1.3 - 22.1 ] & [ 1.4 - 23.6 ]\\
39 & Y & 11.2 & [ 4.6 - 26.3 ] & [ 5.7 - 22.3 ] & [ 2.3 - 52.2 ] & [ 2.2 - 51.0 ]\\
40 & Zr & 11.0 & [ 4.6 - 26.1 ] & [ 6.3 - 19.0 ] & [ 2.6 - 45.3 ] & [ 2.6 - 44.2 ]\\
41 & Nb & 8.0 & [ 3.3 - 18.9 ] & [ 2.1 - 30.4 ] & [ 0.9 - 72.1 ] & [ 0.8 - 65.8 ]\\
42 & Mo & 11.4 & [ 4.9 - 27.3 ] & [ 4.7 - 27.5 ] & [ 2.0 - 65.8 ] & [ 2.3 - 75.5 ]\\
44 & Ru & 2.4 & [ 1.1 - 5.6 ] & [ 0.9 - 6.0 ] & [ 0.4 - 14.2 ] & [ 0.5 - 15.2 ] \\
\enddata
\end{deluxetable}

%

\vspace{5mm}




\bibliographystyle{yahapj}
\bibliography{ms}

\end{document}